\documentclass[a4paper,11pt,twoside]{report}
\usepackage{latexsym,epsfig,amsmath,times}

\newcommand{\hf}{H}
\newcommand{\hfp}{\hf_+}
\newcommand{\hfm}{\hf_-}
\newcommand{\hfin}{\hf_{\text{in}}}
\newcommand{\hfpin}{\hf_{+,\text{in}}}
\newcommand{\hfout}{\hf_{\text{out}}}
\newcommand{\hfpout}{\hf_{+,\text{out}}}
\newcommand{\lf}{l}
\newcommand{\Mp}{M_{\text{p}}}
\newcommand{\op}{\omega_{\text{p}}}
\newcommand{\bra}[1]{\langle#1|}

\newcommand{\ket}[1]{|#1\rangle}
\newcommand{\ketin}[1]{|#1;\text{in}\rangle}
\newcommand{\ketout}[1]{|#1;\text{out}\rangle}
\newcommand{\sprod}[2]{\langle#1|#2\rangle}
\newcommand{\eval}[3]{\langle#1|#2|#3\rangle}
\newcommand{\La}{\mathcal L}
\newcommand{\Lap}{\La_+}
\newcommand{\Lam}{\La_-}
\newcommand{\dphm}{\Delta_+^{\text{HM}}}
\newcommand{\dpnr}{\Delta_+^{\text{NR}}}
\newcommand{\Laphm}{\La_{+,\text{HM}}^0}
\newcommand{\Lapnr}{\La_{+,\text{NR}}^0}
\newcommand{\Dphm}{D_{+,\text{HM}}}
\newcommand{\Dpnr}{D_{+,\text{NR}}}
\newcommand{\gav}[1]{\langle\langle#1\rangle\rangle}

\DeclareMathOperator{\Tr}{Tr}

\begin{document}
\pagestyle{empty}
\begin{titlepage}
  \vspace{2cm}

  \noindent
  {\Huge Matching in Nonrelativistic Effective Quantum Field Theories}
  
  \vspace{11cm}

  \noindent
  Inauguraldissertation \\
  der Philosophisch-naturwissenschaftlichen Fakult\"at \\
  der Universit\"at Bern

  \vspace{1.5cm}

  \noindent
  vorgelegt von

  \vspace{4mm}

  \noindent
  {\Large Alexander Gall}

  \vspace{3mm}

  \noindent
  von Laupen

  \vspace{1.5cm}

  \noindent
  Leiter der Arbeit: \parbox[t]{7cm}{Prof.~J.~Gasser \\
    Institut f\"ur theoretische Physik \\
    Universit\"at Bern}
  
  \cleardoublepage
  \noindent
  \rule{0pt}{4cm}
  \hspace{6cm}
  \parbox[t]{8cm}{
      {\it They began to weave curtains of darkness. \\
      They erected large pillars round the Void,\\
      With golden hooks fastend in the pillars;\\
      With infinite labour the Eternals \\
      A woof wove, and called it Science.} \\ \\
    William Blake, The Book of Urizen}
\end{titlepage}

%%% Local Variables: 
%%% mode: latex
%%% TeX-master: "diss"
%%% End: 
\cleardoublepage
\pagestyle{headings}
\pagenumbering{roman}
\tableofcontents
\pagestyle{myheadings}
\cleardoublepage
\pagenumbering{arabic}

\chapter{Introduction}
%=====================
\markboth{INTRODUCTION}{}
\label{chap:introduction}
Relativistic quantum field theories (RQFTs) describe the interaction of
particles at energies accessible in today's experiments. In most cases, exact
solutions are not known and one has to resort to perturbation theory or
lattice calculations. The former is only valid at energies where the
interaction is small (in the asymptotic region). In an infrared free theory
like QED, there can be an additional problem when the energies and momenta of
the process under consideration become small.  Consider, for example, the
differential cross section of $e^+e^-$ scattering in the center of mass system
to leading order.  Expanding in the relative velocity $v$ one finds that it
diverges like $1/v^4$. The reason for this nonsensical result is the fact that
perturbation theory breaks down at a scale of the order of $m_e\alpha^2$ --
one would have to sum infinitely many graphs that all give contributions of
the same order of magnitude. Non-perturbative effects of this kind are
notoriously difficult to handle in a RQFT.

In this particular process, the particles can form a bound state which shows
up as an isolated pole of the fermionic four-point function in the center of
mass momentum, which cannot be seen to any finite order in ordinary
perturbation theory.  The tool to study this object is the so-called
homogeneous Bethe-Salpeter equation. It is a fourth-order integro-differential
equation for the ``wave function'', which is essentially the residue of the
pole. This is a rather complicated object and no methods are known to solve
this difficult mathematical problem exactly.

All approaches to solve the Bethe-Salpeter equation perturbatively take
advantage of the fact that the scale $m_e\alpha^2$ is much smaller than $m_e$,
suggesting that a non-relativistic approximation is a good starting point. It
turns out that this procedure suffers from numerous technical problems and
despite the long history of the topic, there is to date no truly systematic
perturbation theory available.

Caswell and Lepage~\cite{caswell-lepage} pointed out that the traditional
approach is not well adapted to the non-relativistic nature of the problem.
After all, simple quantum mechanics gives the energy levels of positronium
quite accurately. They recognized that the source of all the problems is the
existence of a hierarchy of physical scales: the electron mass $m_e$, the
typical bound state momentum $m_e\alpha$ and the bound state energy
$m_e\alpha^2$. In the relativistic treatment, all of these scales are present
in the integral kernel of the Bethe-Salpeter equation and it is very difficult
to expand it systematically. They suggest that one should first construct an
effective theory, in which the physics that takes place at scales of the order
of $m_e$ or higher are represented by local interactions of the fields, which
are suppressed by powers of $1/m_e$.  The coefficients of these terms are
determined by comparing scattering amplitudes with those of the full theory at
energies where bound states can be neglected.  With the information about high
energies encoded in the effective couplings, one can then perform bound state
calculations with the effective theory.

The point is that the remaining physical scales are much
smaller than $m_e$. As a consequence, no additional heavy particles (of mass
$m_e$)  can be created\footnote{In positronium, $e^+$ and $e^-$ can, however,
  annihilate into photons. We don't want to go into this rather subtle issue
  here and ignore this effect. To be save, we could consider a stable system,
  like $e^+\mu^-$ as is actual done in~\cite{caswell-lepage}, introducing
  another scale $m_\mu$.} and the theory is confined to a subspace of the Fock
space in which their number is conserved. This is precisely the setting one
has in quantum mechanics: As long as we don't try to resolve processes taking
place at a scale of the order of the Compton wavelength $1/m_e$, the
description in terms of a wave function that obeys a Schr\"odinger equation
is perfectly adequate. 

The concept of a non-relativistic quantum field theory (NRQFT) outlined above
is the bridge between field theory and quantum mechanics. It is equivalent to
the full theory below the heavy scale but takes advantage of the
non-relativistic character of some degrees of freedom by incorporating
relativistic effects in a systematic expansion in inverse powers of some heavy
scale $M$. The interaction of heavy particles is described by a Schr\"odinger
equation whose Hamilton operator is obtained from the Lagrangian of the NRQFT.
What seemed so hard to do in the RQFT, namely the summation of the
non-perturbative part of the theory, simply amounts to solving a lowest
order approximation of this Schr\"odinger equation exactly. One can then use
standard methods of quantum mechanics to perform a systematic perturbation
theory from there.

The formalism has been applied to various processes with considerable success
(the following references are only a selection and by no means complete).
NRQCD, the low-energy version of QCD, was used to study bound states of heavy
quarks by Bodwin, Braaten and Lepage~\cite{lepage-bodwin-braaten}. Muonium and
Positronium hyperfine splitting was already considered
in~\cite{caswell-lepage} and later extended to higher order
corrections~\cite{labelle-thesis,labelle-lepage,labelle-zebarjad-burgess,hoang-labelle-zebarjad,kinoshita-nio-1,kinoshita-nio-2}.

%The bound states still provides two different energy scales $m_e\alpha$ and
%$m_e\alpha^2$ and it turned out to be quite difficult to have a systematic and
%manifest power counting even in the NRQFT. The literature that addresses this
%problem can be traced, for example, from ref.~\cite{griesshammer-1}.

Another system where a NRQFT approach can be useful is the bound state formed
by $\pi^+$ and $\pi^-$.  Because the binding energy is of the order of keV, it
probes the $\pi\pi$ interaction practically at threshold. The decay width of
this atom is related to the $\pi\pi$ scattering lengths and will be measured
soon in the DIRAC experiment at CERN~\cite{DIRAC}, providing a high precision
test of low energy QCD. The leading term of the lifetime was given by Uretsky
and Palfrey~\cite{uretsky-palfrey}. Recently, corrections have been calculated
using different techniques to solve the Bethe-Salpeter equation in the
relativistic
framework~\cite{jallouli-sazdjian,rusetsky1,rusetsky2,rusetsky3,rusetsky4}.
Another approach, based on non-relativistic potential models, was pursued by
the authors of refs.~\cite{gashi,badertscher-1,badertscher-2}.  First attempts
using a NRQFT approach have been
published~\cite{labelle-buckley,kong-ravndall} but need further clarification.

Let us also mention that there is a different branch of NRQFTs, where there is
only one heavy particle involved. In this case, the scales $m_e\alpha$,
$m_e\alpha^2$ are absent and with them the non-perturbative effects. The heavy
particle can be considered to be static and power counting becomes very
simple. This version of a NRQFT is used for the description of mesons
containing one heavy quark under the name of heavy quark effective theory
(HQET) and also for the pion-nucleon system where it is called heavy baryon
chiral perturbation theory (HBCHPT). See
refs.~\cite{hqet-review,hbchpt-review} for reviews on these subjects.

A crucial step in the construction of a NRQFT is the matching with the
fundamental theory, where the coupling constants are adjusted such that the
scattering amplitudes agree to some order in inverse powers of the heavy
scale. In order to do this, one needs to renormalize both theories, i.e.
introduce a regularization scheme that allows to absorb the divergences of
Green's functions into the coupling constants in a systematic way. Also, one
has to express the physical mass of the heavy particle in terms of the
parameters of the theory and determine the effective normalization of the
field (the ``wave function renormalization'') due to self energy effects. This
is certainly no problem in the RQFT, which is expressed in a Lorentz covariant
form. The NRQFT is not covariant and it is not a priori clear how these tasks
should be performed there. The fact that it took some time for people to 
realize that in some versions of HQET and HBCHPT the fields were incorrectly
normalized even at tree level~\cite{balk:1993,hbchpt-wave-function-ecker} shows
that this question is not as innocent as it may seem. Unfortunately, the
discussions are often obscured by the formalism of the particular model under
consideration.

However, as in a RQFT, the procedure of mass and wave function renormalization
is independent of a particular model and can be treated once and for all in
the language of the one-particle irreducible two-point
function. To the best of the author's knowledge, such a discussion is not
available in the literature. The present work tries to fill this gap by
studying how this mechanism works in the case of a heavy scalar field. 
We only consider Yukawa-type couplings to
other scalar fields to avoid complications due to gauge symmetry and spin.

This work is organized as follows. In chapter~\ref{chap:matching} we show how
amplitudes and Green's functions of a generic Lagrangian with one heavy scalar
field can be matched with the corresponding effective theory. In
chapter~\ref{chap:toy-model}, we consider a toy-model and explicitly construct
two non-local effective Lagrangians that are equivalent to the full theory in
the pure particle- and anti-particle sectors, verifying the general statements
made in chapter~\ref{chap:matching}.  Finally, the $1/M$ expansion of
tree-level Green's functions and Amplitudes in the full theory is discussed in
chapter~\ref{chap:M-expansion} and it is shown how they can be reproduced by
the effective theory order by order in powers of the inverse heavy scale.

%%% Local Variables: 
%%% mode: latex
%%% TeX-master: "diss"
%%% End: 
\chapter{Matching in the Particle Sector}
%========================================
\markboth{MATCHING IN THE PARTICLE SECTOR}{}
\label{chap:matching}

\section{Transition Amplitudes}
%------------------------------
\label{sec:transition-amplitudes}
To have a specific example and to keep things simple at the same time, we
consider a theory of the form
\begin{align}
  \label{eq:L-gener}
  \La &= \La^0+ \bar\La^0 + \La^{\text{int}} \notag \\
  \La^0 &= \partial_\mu\hf^*\partial^\mu\hf-M^2\hf^*\hf.
\end{align}
Here, $\bar\La^0$ contains the kinetic part of all the fields that interact
with $\hf$ through the interaction Lagrangian $\La^{\text{int}}$. We assume
that the masses of these fields are all much smaller than $M$, i.e. $\hf$ is
the only heavy degree of freedom. As such, they appear unaltered in the
effective theory that describes physics at a scale much smaller than $M$.
Therefore, we first concentrate on processes among heavy particles alone.

The free Lagrangian of the heavy field has a $U(1)$ symmetry and the particles
carry a charge which is conserved in all processes if $\La^{\text{int}}$
respects this symmetry. We shall refer to the two types of field quanta as
particle- and anti-particle. They enter the free Lagrangian symmetrically and
can only be distinguished by the interaction with an external field.
Scattering processes which are related by crossing are described by the same
invariant amplitude.

\subsection{Relativistic Theory}
%-------------------------------
\label{sec:full-theory}
The fundamental objects we have to study are the connected Green's functions
\begin{equation}
  \label{eq:G-conn}
  G^{(2n)}(x,y) = \eval{0}{T\hat\hf(x)\hat\hf^\dag(y)}{0}_c.
\end{equation}
Here, $x,y$ are vectors $(x_1,\dots,x_n)$, $(y_1,\dots,y_n)$ and we use the
notation $\hat f(x)\equiv f(x_1)\dots f(x_n)$. Further notation is given in
appendix~\ref{app:notation}.  To each external momentum corresponds a
two-point function $G^{(2)}$ and we define the truncated function
$G_{tr}^{(2n)}$ by
\begin{equation}
  \label{eq:def-truncation}
  G^{(2n)}(p,q) = \hat G^{(2)}(p)\hat G^{(2)}(q)G_{tr}^{(2n)}(p,q).
\end{equation}
Each of the factors $G^{(2)}(p_i)$ has a pole when the momentum is on the mass
shell $p_i^2=\Mp^2$, where $\Mp$ is the physical mass of the particle. The
scattering amplitude, involving $2n$ heavy particles in this case, is related
to the residue of the multiple pole when all momenta are put on their mass
shells. The precise relation is given by the LSZ formalism summarized in
appendix~\ref{app:reduction} for the case at hand. Applied to the process
with $n$ heavy particles in the initial and final states
\begin{multline}
  \sprod{p_1,\dots,p_n;\text{out}}{q_1,\dots,q_n;\text{in}}
  = \sprod{p_1,\dots,p_n;\text{in}}{q_1,\dots,q_n;\text{in}} \\
  + i(2\pi)^4\delta^4\left(P-Q\right) T_{n\rightarrow n},
\end{multline}
where $P=\sum_{i=1}^n p_i$ and $Q=\sum_{i=1}^n q_i$, we find
\begin{equation}
  T_{n\rightarrow n} = \frac{1}{i}Z_\hf^n
  \left.G_{tr}^{(2n)}(p,q)\right|_{\text{on-shell}}.
\end{equation}
``On-shell'' means $p_i^0=\op(\mathbf p_i) = \sqrt{\Mp^2+\mathbf p_i^2},
q_i^0=\op(\mathbf q_i)$ and $Z_\hf$ is the residue of the two-point function
$G^{(2)}$. Note that, due to the manifest covariance of the theory, this
quantity transforms as a scalar under the Lorentz group.

In such a process, heavy anti-particles are only involved as virtual states.
Therefore, it should be possible to remove them as an explicit degree of
freedom and incorporate them into the interaction.

\subsection{Separating Particles and Anti-Particles}
%---------------------------------------------------
\label{sec:free-fields}
The first step towards this goal is to separate particles and anti-particles
in the free field. Consider the equation of motion
\begin{equation}
  \label{eq:KG-eq}
  (\Box + M^2)\hf = 0
\end{equation}
obtained from $\La^0$. The most general solution is a superposition of plane
waves
\begin{equation}
  \hf(x) = \int \frac{d^3p}{(2\pi)^32\omega(\mathbf
  p)}\left(a(p)e^{-ipx}+b^*(p)e^{ipx}\right).
\end{equation}
To separate the positive and negative frequency contributions, we define the
differential operators (see also appendix~\ref{app:KG})
\begin{align}
  D_\pm &= \pm i\partial_t - \sqrt{M^2-\Delta} \\
  d &= (2\sqrt{M^2-\Delta})^{-\frac{1}{2}}
\end{align}
and set
\begin{equation}
  \label{eq:def-hpm}
  \hf_\pm = -D_\mp d\hf.
\end{equation}
With this choice we have
\begin{align}
  \hfp(x) &= \int\frac{d^3p}{(2\pi)^3\sqrt{2\omega(\mathbf p)}} a(p) e^{-ipx}
  \\
  \hfm(x) &= \int\frac{d^3p}{(2\pi)^3\sqrt{2\omega(\mathbf p)}} b^*(p) e^{ipx}
\end{align}
and
\begin{equation}
  \hf = d(\hfp + \hfm).
\end{equation}
The operator $d$ is not really necessary for this decomposition and was only
introduced for later convenience. The fields $\hf_\pm$ satisfy the equations
\begin{equation}
  D_\pm\hf_\pm(x) = 0,
\end{equation}
which are the Euler-Lagrange equations of the Lagrangians
\begin{equation}
  \La_\pm^0 = \hf_\pm^*D_\pm\hf_\pm.
\end{equation}
After canonical quantization, the operators $\hfp^\dag$ and $\hfm$ create a
particle and an anti-particle state, respectively (see
appendix~\ref{app:canonical-quantisation}).

\subsection{Effective Theory in the Particle Sector}
%---------------------------------------------------
\label{sec:eff-particle-theory}
A theory for the particle sector should therefore be of the form
\begin{equation}
  \label{eq:Lp-gener}
  \Lap = \Lap^0 + \bar\La^0 + \Lap^{\text{int}}. 
\end{equation}
In appendix~\ref{app:reduction} it is shown that the Fock space of the free
heavy particles, in which the incoming and outgoing particles live, is the
same as in the relativistic theory. This is obviously a necessary condition
for the existence of an interpolating field that should reproduce transition
amplitudes of a relativistic theory.

The interaction Lagrangian is a local function of the fields and their
derivatives and can be written as
\begin{equation}
  \Lap^{\text{int}} = \sum_{\nu=1}^\infty \frac{1}{M^\nu}\Lap^\nu,
\end{equation}
where $\Lap^\nu$ contains $\nu$ space or time derivatives.  This means that we
deal here with an effective field theory in which $M$ is considered to be a
hard scale. It can only describe processes in which all relevant scales are
much smaller than that. In practice, one always truncates the Lagrangian at
some power in $1/M$ but for the sake of the following arguments, let us assume
that we have summed up the contributions to all orders and postpone the
discussion of this issue.

$U(1)$ symmetry of the Lagrangian insures that the heavy field only occurs in
the combination $\hfp^\dag\hfp$, which means that the number of heavy particles
is conserved at each vertex ($\hfp$ destroys an incoming particle and
$\hfp^\dag$ creates an outgoing one) and therefore for any process (this is
simply the consequence of charge conservation when there is only one type of
charge). The theory is thus naturally confined to a subspace of the Fock space
in which the number of heavy particles is fixed.

We can start by writing down the most general interaction Lagrangian which
respects the symmetries of $\La$.  However, we can immediately see that
Lorentz symmetry is already violated by $\Lap^0$.  The question is then, how
much of this symmetry we have to incorporate into $\Lap$ to be able to
calculate a transition amplitude with the correct transformation properties
under the Lorentz group. Let us formulate a pragmatic approach to the problem.

Due to the lack of knowledge of the transformation properties of $\hfp$ under
the Lorentz group\footnote{In HQET and HBCHPT one introduces a four-velocity
  $v_\mu$ to write down the Lagrangian in different frames of reference and
  Lorentz invariance is replaced by ``reparametrisation
  invariance''~\cite{reparametrization-invariance}}, we only require
rotational invariance of the Lagrangian. We can then calculate the connected
Green's functions
\begin{equation}
  G_+^{(2n)}(x,y) = \eval{0}{T\hat\hfp(x)\hat\hfp^\dag(y)}{0}_c.
\end{equation}
Next, we can try to derive a reduction formula for this theory, relating
transition amplitudes to poles of these Green's functions. As shown in
appendix~\ref{app:reduction}, this involves one non-trivial assumption about
the structure of the two-point function, namely that it permits the definition
of a physical mass $\Mp$ so that
\begin{equation}
  G_+^{(2)}(p) = \frac{1}{i}\frac{Z_+(\mathbf p^2)}{\op(\mathbf
  p)-p^0-i\epsilon} + \dots 
\end{equation}
This implies that not all of the coupling constants of the original Lagrangian
are independent. No additional assumptions are needed to define the object
\begin{equation}
  T_{n\rightarrow n}^+ = \frac{1}{i}\prod_{i=1}^nZ_+(\mathbf p_i^2)^\frac{1}{2}
    Z_+(\mathbf 
    q_i^2)^\frac{1}{2} 
  \left.G_{+,tr}^{(2n)}(p,q)\right|_{\text{on-shell}},
\end{equation}
where the truncated function is defined by
\begin{equation}
  G_+^{(2n)}(p,q) = \hat G_+^{(2)}(p)\hat G_+^{(2)}(q)G_{tr}^{(2n)}(p,q).
\end{equation}
$T_{n\rightarrow n}^+$ does not yet transform as a scalar under the Lorentz
group as it should if it is supposed to reproduce $T_{n\rightarrow n}$.

\subsection{Matching}
%--------------------
\label{sec:amplitude-matching}
Symmetry only fixes each term in the Lagrangian up to a factor. These low
energy constants (LEC) are at our disposal and can be chosen in such a way
that all scattering amplitudes considered above are identical. This procedure
is called matching. Before we formulate it, we should say a word about the
normalization of one-particle states, because the amplitudes clearly depend on
them. Although arbitrary, there is still a most natural choice of
normalization (see also appendix~\ref{app:canonical-quantisation}). In the
full theory, we chose it to be Lorentz invariant
\begin{equation}
  \sprod{p}{p'} = \sprod{\bar p}{\bar p'} = 2\op(\mathbf
  p) (2\pi)^3 \delta^3(\mathbf p-\mathbf p'),  
\end{equation}
where as for $\Lap$ we chose 
\begin{equation}
  \sprod{p}{p'} =(2\pi)^3 \delta^3(\mathbf p-\mathbf p').  
\end{equation}
Therefore, before we try to match amplitudes, we must make up for this
difference in normalization by replacing, say, the states used in the
effective theory by
\begin{equation}
  \ket{p} \rightarrow \sqrt{2\omega_{\text{p}}(\mathbf p)}\ket{p}.
\end{equation}
The matching condition then reads
\begin{equation}
  \label{eq:T-matching}
  T_{n\rightarrow n} = \prod_{i=1}^n\sqrt{2\op(\mathbf p_i)}
    \sqrt{2\op(\mathbf q_i)} T_{n\rightarrow n}^+,
\end{equation}
which automatically restores Lorentz symmetry for the transition
amplitudes. In the last section we have seen that the effective theory is
actually an expansion in inverse powers of the heavy scale $M$. The matching
can only make sense if the relativistic amplitude possesses such an expansion
in the region of phase space we are interested in.

%%% Local Variables: 
%%% mode: latex
%%% TeX-master: "~/diss/tex/diss"
%%% End: 

\section{Green's Functions}
%--------------------------
\label{sec:greens-functions}
The matching of scattering amplitudes involves only Green's functions
evaluated on the mass shell of all particles involved.  They are, however,
also interesting in the unphysical region because they reflect general
properties of quantum field theories like unitarity in their non-trivial
analytic structure. It is interesting to see how the Green's functions of the
fundamental theory compare to the ones of the effective theory.

Being unphysical quantities, off-shell Green's functions have no unique
definition. Redefinitions of the fields that do not change the classical field
theory give, in general, different off-shell results while describing the same
physics. Suppose we have chosen a particular off-shell extrapolation in the
fundamental theory. Naively, one may be tempted to identify the truncated
functions $G_{tr}^{(2n)}$ with $G_{+,tr}^{(2n)}$, i.e. consider the latter to
be the $1/M$ expansion of the former. One would then expect that they differ
only by a polynomial in the momenta which can be absorbed by a proper choice
of coupling constants in the effective theory. However, this is not true, as
we will show now.

Two remarks about the following statements are in order. First, we suppose
that renormalization was performed in both theories and that everything is
finite and well defined.  Second, as mentioned before, the effective theory is
an expansion in $1/M$. Therefore, the matching is actually performed order by
order in $1/M$ and we assume that the relativistic Green's functions can be
expanded in this way.

Let us start with eq.~\eqref{eq:T-matching}. It can be written in terms of the
truncated Green's functions as
\begin{multline}
  \label{eq:T-matching-2}
  Z_\hf^n
  \left.G_{tr}^{(2n)}(p,q)\right|_{\text{on-shell}}= \\
  \prod_{i=1}^n (Z_+(\mathbf p_i^2)\op(\mathbf p_i))^\frac{1}{2} (Z_+(\mathbf
  q_i^2)\op(\mathbf p_i))^\frac{1}{2}
  \left.G_{+,tr}^{(2n)}(p,q)\right|_{\text{on-shell}}.
\end{multline}
Without knowing the relationship between the residues $Z_\hf$ and $Z_+$, we
cannot express, say, $G_{tr}^{(2n)}$ in terms of quantities that can be
calculated with $\Lap$ alone. In appendix~\ref{app:2pf} it is shown how such a
relation emerges from the matching of the two-point functions. The statement
is that when the irreducible parts $\Sigma$,~$\Sigma_+$ defined by
\begin{align}
  G^{(2)}(p) &= \frac{1}{i}\frac{1}{M^2-p^2 + i\Sigma(p^2) -i\epsilon} \\
  G_+^{(2)}(p) &= \frac{1}{i}\frac{1}{\omega(\mathbf p)-p^0 +
    i\Sigma_+(p^0,\mathbf p^2) - i\epsilon},
\end{align}
are matched according to
\begin{equation}
  \Sigma_+(p^0,\mathbf p^2) = \frac{\Sigma(p^2)}{2\omega(\mathbf p)+
    \frac{i\Sigma(p^2)}{\omega(\mathbf p)+p^0}},
\end{equation}
the physical masses defined by
\begin{align}
  \Mp &= M^2 + i\Sigma(\Mp^2) \\
  \op(\mathbf p) &= \sqrt{\Mp^2+\mathbf p^2} = \omega(\mathbf p) +
  i\Sigma_+(\op(\mathbf p), \mathbf p^2)
\end{align}
are identical and the residues of
\begin{align}
  G^{(2)}(p) &= \frac{1}{i}\frac{Z_\hf}{\Mp^2-p^2-i\epsilon} +
  \text{regular},p^2\rightarrow\Mp^2 \\
  G_+^{(2)}(p) &= \frac{1}{i}\frac{Z_+(\mathbf p^2)} {\op(\mathbf
    p)-p^0-i\epsilon} + \text{regular},p^0\rightarrow\op(\mathbf p)
\end{align}
are related by
\begin{equation}
  Z_+(\mathbf p^2) = \frac{(\omega(\mathbf p)
  + \op(\mathbf p))^2}{4\op(\mathbf p)\omega(\mathbf p)} Z_\hf.
\end{equation}
If we plug this into eq.~\eqref{eq:T-matching-2}, we find
\begin{equation}
  \label{eq:T-matching-3}
  \left. G^{(2n)}_{tr}(p,q)\right|_{\text{on-shell}} = \prod_{i=1}^n
  \frac{\omega(\mathbf p_i)+\op(\mathbf 
  p_i)}{\sqrt{2\omega(\mathbf p_i)}} \frac{\omega(\mathbf q_i)+\op(\mathbf
  q_i)}{\sqrt{2\omega(\mathbf q_i)}}
  \left.G_{+,tr}^{(2n)}(p,q)\right|_{\text{on-shell}}
\end{equation}
and all quantities on the r.h.s. can be calculated with the Lagrangian $\Lap$.
Let us extend this relation to off-shell Green's functions. For this purpose
we define a new truncation procedure
\begin{equation}
  \label{eq:def-trunc-bar}
  G_+^{(2n)}(p,q) = \hat \mathcal G_+(p)
  \hat \mathcal G_+(q) \bar G_{+,tr}^{(2n)}(p,q)
\end{equation}
with
\begin{equation}
  \label{eq:eff-ext-line}
    \mathcal G_+(p) \doteq 
  \frac{G_+^{(2)}(p)}{\sqrt{2\omega}}
  \left(1-\frac{i\Sigma_+(p^0,\mathbf p^2)}{\omega(\mathbf p) + p^0}\right)
\end{equation}
and impose the off-shell matching condition
\begin{equation}
  \label{eq:gf-matching}
  G_{tr}^{(2n)}(p,q) = \bar G_{+,tr}^{(2n)}(p,q),
\end{equation}
which indeed reduces to eq~\eqref{eq:T-matching-3} on the mass shell. The
functions $G_{+,tr}^{(2n)}$ and $\bar G_{+,tr}^{(2n)}$ differ essentially by
the self-energy $\Sigma_+$, which is a non-trivial function of momentum. This
is the reason why a matching between the ``naturally'' truncated functions
$G_{tr}^{(2n)}$ and $G_{+,tr}^{(2n)}$ is impossible - they differ by more than
just a polynomial. This point will be illustrated in
section~\ref{sec:p-a-sector} in a simple toy-model.

%%% Local Variables: 
%%% mode: latex
%%% TeX-master: "~/diss/tex/diss"
%%% End: 
\chapter{Construction of the Effective Lagrangian for a simple Model}
%===================================================================
\label{chap:toy-model}
\markboth{CONSTRUCTION OF THE LAGRANGIAN FOR A SIMPLE MODEL}{}
\section{The Model}
%-----------------
The model we are considering is given by
\begin{align}
  \label{eq:L-toy-model}
  \bar\La^0 &\equiv \La_\lf^0 =  \frac{1}{2}\partial_\mu \lf\partial^\mu
  \lf-\frac{m^2}{2}\lf^2 \notag \\
  \La^{\text{int}} &= e\hf^*\hf \lf
\end{align}
in the notation of section~\ref{sec:transition-amplitudes}. To stay in the
scope of that section we chose $m\ll M$ and refer to $l$ as the light field. It
will always keep its relativistic form. In the following, we will explicitly
construct an effective theory of the form given in eq.~\eqref{eq:Lp-gener}
that can be proven to reproduce the scattering amplitudes in the sector where
there is a fixed number of heavy particles and an arbitrary number of light
particles.

%%% local variables: 
%%% mode: latex
%%% TeX-master: "~/diss/tex/diss"
%%% End: 
\section{Interaction with an External Field}
%-------------------------------------------
\label{sec:external-field}
In this section, the light field $\lf$ is a given function of space and time
and we consider the Lagrangian
\begin{equation}
  \label{eq:L-h-l-ext}
  \La^{ext} = \La^0 + \La^{\text{int}}+ j^*\hf + \hf^*j.
\end{equation}
The equation of motion
\begin{equation}
  \label{eq:eom-h-l-ext}
  D_e \hf \doteq (D_M-e\lf)\hf = j,
\end{equation}
where $D_M = \Box+M^2$, has the formal solution
\begin{equation}
  \label{eq:sol-h-l-ext}
  \hf = D_e^{-1}j.
\end{equation}
$D_e^{-1}$ is the complete two-point function of this theory and can be
expressed in terms of the free propagator $D_M^{-1}$ defined in
appendix~\ref{app:KG} as
\begin{equation}
  \label{eq:inv-De-as-inv-D}
  D_e^{-1} = D_M^{-1}\frac{1}{1-e\lf D_M^{-1}}.
\end{equation}
We define the truncated two-point function $T$ by
\begin{equation}
  \label{eq:def-T}
  D_e^{-1} = D_M^{-1} + D_M^{-1}TD_M^{-1}.
\end{equation}
In perturbation theory, it is simply a string of free propagators with
insertions of the external field
\begin{equation}
  \label{eq:T-perturbative}
  T = e\lf + e^2 \lf D_M^{-1}\lf+O(e^3).
\end{equation}
All information about a particle moving in the external field is contained in
this operator. The possible physical processes are the scattering of a
particle or an anti-particle (including the formation of bound states if the
external field allows them), pair-annihilation and pair-creation. We are about
to construct two independent non-local theories that can reproduce the
scattering processes for particles and anti-particles separately. To this end,
we define the fields $\hf_\pm$ as in eq.~\eqref{eq:def-hpm} and introduce the
vectors
\begin{align}
  \vec\hf &=
  \begin{pmatrix}
    \hfp \\
    \hfm
  \end{pmatrix} &
  \vec j &=
  \begin{pmatrix}
    j \\
    j
  \end{pmatrix}
\end{align}
and the operator
\begin{align}
  D &=
  \begin{pmatrix} 
    A & eB \\
    eB & C
  \end{pmatrix}
  \notag \\
  A &= D_+ + eB \notag \\
  C &= D_- + eB \notag \\
  B &= d\lf d.
\end{align}
It is easy to check that $\vec\hf$ obeys
\begin{equation}
  \label{eq:eom-H-l-ext}
  D \vec\hf = -d\vec j. 
\end{equation}
Writing $D^{-1}$ as
\begin{equation}
  \label{eq:def-inf-DFW}
  D^{-1} = 
  \begin{pmatrix} 
    G_1 & G_2 \\
    G_3 & G_4
  \end{pmatrix}
\end{equation}
and using the fact that $j$ is arbitrary, we find the operator identity
\begin{equation}
  \label{eq:De-decomposition}
  D_e^{-1} = -\sum_{n=1}^4 d G_n d.
\end{equation}
To explore the significance of this, we investigate the structure of the
$G_n$. They can be expressed in terms of the $A,B,C$ defined above by solving
the equation $D D^{-1} = \mathbf 1$. Their structure in terms of the Green's
functions $D_\pm^{-1}$ allows for a definition of truncated objects
$T_{\pm\pm}$ just like in (\ref{eq:def-T})
\begin{align}
  \label{eq:def-G1}
  G_1 &= (A-e^2BC^{-1}B)^{-1} \doteq D_+^{-1} - D_+^{-1}T_{++}D_+^{-1} \\
  G_2 &= -eA^{-1}BG_4 \doteq -D_+^{-1}T_{+-}D_-^{-1} \\
  G_3 &= -eC^{-1}BG_1 \doteq -D_-^{-1}T_{-+}D_+^{-1} \\
  G_4 &= (C-e^2BA^{-1}B)^{-1} \doteq D_-^{-1} - D_-^{-1}T_{--}D_-^{-1}.
\end{align}
It is straight forward to show that the $T_{\pm\pm}$ are all essentially
equal to $T$ (see appendix~\ref{app:T}). More precisely, we find that
\begin{equation}
  \label{eq:Tpm-T-equivalence}
  T_{++}=T_{+-}=T_{-+}=T_{--}=dTd
\end{equation}
holds to all orders in perturbation theory. We have therefore found a
decomposition of the r.h.s. of eq.~\eqref{eq:def-T} in which each of the four
pieces contains the complete truncated function $T$.  We define a non-local
Lagrangian for each of the fields $\hf_\pm$ by
\begin{eqnarray}
  \label{eq:L-hp-ext}
  \mathcal L_\pm^{ext} &=& \hf_\pm^*\mathcal D_\pm\hf_\pm \nonumber \\
  \mathcal D_+ &=& A -e^2BC^{-1}B \nonumber \\
  \mathcal D_- &=& C -e^2BA^{-1}B.
\end{eqnarray}
The associated two-point functions
\begin{align}
  \eval{0}{T\hfp(x)\hfp^\dag(y)}{0} &= iG_1(x,y) \\
  \eval{0}{T\hfm(x)\hfm^\dag(y)}{0} &= iG_4(x,y)
\end{align}
contain all the information about the interaction of one particle and one
anti-particle with the external field, respectively. Note that pair creation
or annihilation processes are not included: the fields $\hf_\pm$ do not talk
to each other.

Let us illustrate the connection between the original Lagrangian and these two
effective Lagrangians for the case of scattering in a static field
$\lf=\lf(\mathbf x)$. In the notation for in- and out states introduced in
appendix~\ref{app:reduction}, the transition amplitudes $T_\pm$ for particle-
and anti-particle scattering are defined by
\begin{align}
  \sprod{p;\text{out}}{q;\text{in}}
  &= \sprod{p;\text{in}}{q;\text{in}}+i2\pi\delta(p^0-q^0) T_+(p,q) \\
  \sprod{\bar p;\text{out}}{\bar q;\text{in}} &= \sprod{\bar p;\text{in}}{\bar
    q;\text{in}}+i2\pi\delta(p^0-q^0) T_-(p,q).
\end{align}
Fourier transformation is defined as in appendix~\ref{app:notation} with the
difference that only the energy is conserved
\begin{equation}
  2\pi\delta(p^0-q^0)T(p,q) = \int d^4x d^4y e^{i(px-qy)}T(x,y).
\end{equation}
The physical momenta of incoming particles (anti-particles) and outgoing
particles (anti-particles) are given by $q(-q)$ and $p(-p)$, respectively.
Applying the reduction formula of appendix~\ref{app:reduction}, we find in the
full theory
\begin{equation}
  T_\pm(p,q) = \left. T(\pm p,\pm q)\right|_{p^0=
  q^0=\omega(\mathbf p)},
\end{equation}
whereas the effective theories give
\begin{equation}
  T_\pm(p,q) = \left. \frac{1}{\sqrt{2\omega(\mathbf p)}}
  \frac{1}{\sqrt{2\omega(\mathbf q)}} T(\pm p,\pm q)\right|_{p^0= 
  q^0=\omega(\mathbf p)}.
\end{equation}
The additional kinematical factors $1/\sqrt{2\omega}$ are due to the different
normalizations of free one-particle states. We have thus verified that the
Lagrangians $\La_\pm^{ext}$ produce scattering amplitudes that automatically
satisfy the matching condition stated in eq.~\eqref{eq:T-matching}.

%%% Local Variables: 
%%% mode: latex
%%% TeX-master: "~/diss/tex/diss"
%%% End: 
\section{Non-local Lagrangians in the Particle and Anti-\-Par\-ticle Sector}
%-----------------------------------------------------------------------
\label{sec:p-a-sector}
We return to the original Lagrangian defined in eq. \eqref{eq:L-toy-model},
where $\lf$ represents a dynamical degree of freedom. The results of the last
section can be used to construct two non-local Lagrangians that are equivalent
to the original theory in the pure particle- and anti-particle sectors of the
heavy field including any number of light fields. Ultimately, these
Lagrangians will be brought to a local form by expanding in $1/M$. It is the
expanded version that is a true {\it effective} theory in the sense that it
reproduces the fundamental theory only at low energies. The non-local version
still contains the complete information about truncated Green's functions as
we are about to show now.

\subsection{Green's Functions}
%-----------------------------
\label{sec:p-q-greens-functions}
We consider the generating functional $Z$ of all Green's functions and perform
the integration over the heavy field. In appendix \ref{app:det} it is shown
that it can be written in the form
\begin{align}
  \label{eq:Z-integrated}
  Z[j,j^*,J] &= \frac{1}{\mathcal Z} \int [d\lf](\det D_+^{-1}\mathcal
  D_+)^{-1}e^{i\int
    \mathcal L_\lf^0 + j^*D_e^{-1}j+J\lf} \\
  \mathcal Z &= \int [d\lf](\det D_+^{-1}\mathcal D_+)^{-1}e^{i\int \mathcal
    L_\lf^0},
\end{align}
with $D_e$ and $\mathcal D_+$ given in \eqref{eq:eom-h-l-ext} and
\eqref{eq:L-hp-ext}, respectively. The determinants are evaluated in $D\neq 4$
dimensions where they are finite to all orders in perturbation theory, i.e. we
deal here with a regularized but not renormalized theory. The statements
derived in this section are a priori only valid within this framework. In
appendix~\ref{app:1-loop-renormalization}, we determine the counter terms
necessary to render all Green's functions finite in $D=4$ to one loop (i.e.
$O(e^2)$). By working only to this order in perturbation theory, the results
of this section can be proven to hold also in $D=4$.

Now consider the theory defined by
\begin{equation}
  \label{eq:L-hp}
  \mathcal L_+ = \hf_+^*\mathcal D_+\hf_+ + \mathcal L_\lf^0.
\end{equation}
Its generating functional after integration over $\hf_+$ is
\begin{equation}
  \label{eq:Z-hp-integrated}
  Z_+[j,j^*,J] = \frac{1}{\mathcal Z} \int [d\lf] (\det D_+^{-1}\mathcal
  D_+)^{-1} e^{i\int 
  \mathcal L_\lf^0 - j^*\mathcal 
  D_+^{-1}j + J\lf}.
\end{equation}
This is simply $Z$ with $D_e^{-1}$ replaced by $-\mathcal D_+^{-1}$.  In the
last section we have found that they can be written as
\begin{align}
  \label{eq:p-a-T-def}
  D_e^{-1} &= D_M^{-1}\left( 1 + T D_M^{-1}\right) \\
  \label{eq:p-a-D-eff}
  \mathcal D_+^{-1} &= D_+^{-1}\left(1 - dTd D_+^{-1}\right).
\end{align}
The first equation is the definition of $T$ which is to be considered as a
functional of $\lf$ within the path integrals above.  Let us first consider
the $n$-point functions (the tilde distinguishes them from the connected
functions defined below)
\begin{align}
  \label{eq:orig-n-point-f}
  \tilde G^{(a,b)}(x,y,z) &= \langle
  0|T\,\hat\hf(x)\hat\hf^+(y)\hat\lf(z)|0\rangle
  \nonumber \\
  &= \left. \frac{1}{i^n}\frac{\delta^n Z}{\widehat{\delta j}^*(x)
      \widehat{\delta j}(y)\widehat{\delta J}(z)}\right|_{j=j^*=J=0},
\end{align}
where $(a,b)$ is a pair of integers with $2a+b=n$ and $x,y,z$ are vectors
$(x_1,\dots,x_a)$, $(y_1,\dots,y_a)$, $(z_1,\dots,z_b)$. We recall that we use
the shorthand notation for the product of fields and the definition of the
Fourier transform as given in appendix~\ref{app:notation}. The functions
$\tilde G_+^{(a,b)}$ of the effective theory are defined through $Z_+$ in an
analogous manner.

The derivatives with respect to the sources $j$,$j^*$ bring down factors of
$D_e^{-1}$ and $\mathcal D_+^{-1}$ in $Z$ and $Z_+$, respectively. It is clear
that the free parts $D_M^{-1}$ and $D_+^{-1}$ of eqns.~\eqref{eq:p-a-T-def}
and~\eqref{eq:p-a-D-eff} only contribute to disconnected Green's functions
(except for the two-point functions, see below) and we ignore them for the
moment. Denoting a permutation $P$ of the coordinates $y_i$ by
\begin{displaymath}
  P(y_1,\dots,y_a) = (y_{P_1},\dots,y_{P_a}),
\end{displaymath}
the remaining contributions to $\tilde G^{(a,b)}$ and $\tilde
G_+^{(a,b)}$ can then be written as the sum over all permutations of the
term 
\begin{equation}
  \frac{1}{i^a}\frac{1}{\mathcal Z} \int [d\lf](\det D_+^{-1}\mathcal
  D_+)^{-1}\prod_{i=1}^a
  f(x_i,y_{P_i})\prod_{j=1}^b
  \lf(z_j)e^{i\int \La_\lf^0}.
\end{equation}
For $Z$, the function $f$ is given by
\begin{equation}
  f(u,v) = \int d^Ds d^Dt \Delta_M(u-s)T(s,t)\Delta_M(t-v)
\end{equation}
and for $Z_+$ by
\begin{equation}
  f(u,v) = -\int d^Ds d^Dt \Delta_+(u-s)d_s T(s,t) d_t
  \Delta_+(t-v). 
\end{equation}
The point is that $\lf$ only occurs in $T$, which is the same in both
expressions. The free propagators, which form the endpoints of external legs
corresponding to heavy particles, and the differential operators $d$ can be
taken out of the remaining path integral. Since we have already discarded some
disconnected pieces, it is useful to consider only connected Green's functions
denoted by $G^{(a,b)}$ and $G_+^{(a,b)}$, generated by the
functionals $iW$ and $iW_+$ defined by
\begin{align}
  e^{iW[j,j^*,J]} &\doteq Z[j,j^*,J] \\
  e^{iW_+[j,j^*,J]} &\doteq Z_+[j,j^*,J]
\end{align}
in analogy with eq.~\eqref{eq:orig-n-point-f}. What we have found above is
that these functions differ only by the outermost parts of their external
heavy lines. More precisely, if we write ($u$,$v$,$w$ are vectors like
$x$,$y$,$z$ and $\Delta_m$ is the propagator of the light field obtained from
$\Delta_M$ by replacing $M$ by $m$)
\begin{align}
  G^{(a,b)}(x,y,z) =&\; \frac{1}{i^{n}}\int d^Du d^Dv d^D w
  \prod_{i=1}^a\prod_{j=1}^b \Delta_M(x_i-u_i) \\
  & S(u,v,w)\Delta_M(v_i-y_i)\Delta_m(z_j-w_j) \\
  G_+^{(a,b)}(x,y,z) =&\; \frac{(-1)^{2a}}{i^{n}}\int d^Du d^Dv d^D w
  \prod_{i=1}^a\prod_{j=1}^b \Delta_+(x_i-u_i) \\
  & d_{u_i}S_+(u,v,w)d_{v_i}\Delta_+(v_i-y_i)\Delta_m(z_j-w_j),
\end{align}
for $2a+b>2$ we have $S=S_+$ to any order in perturbation theory. In
particular, $G_+^{(a,b)}$ has the full loop structure of $G^{(a,b)}$.

Let us consider the two-point functions of the heavy fields in detail. In
momentum space we find
\begin{align}
  \label{eq:G2-orig}
  G^{(1,0)}(p) &= \frac{1}{i}\Delta_M(p)\left( 1 + S(p)\frac{1}{i}\Delta_M(p)
  \right) \\
  \label{eq:G2-eff}
  G_+^{(1,0)}(p) &= i\Delta_+(p)\left( 1 + \frac{S(p)}{2\omega(\mathbf p)}
    i\Delta_+(p) \right)
\end{align}
with $S(p)$ being the Fourier transform of
\begin{equation}
  S(x-y) = \frac{i}{\mathcal Z}\int [dl](\det D_+^{-1}\mathcal D_+)^{-1}
  T(x,y) e^{i\int \La_\lf^0}.
\end{equation}
The interesting thing about this is that the irreducible two-point functions
$\Sigma$, $\Sigma_+$ defined by
\begin{align}
  G^{(1,0)}(p) &= \frac{1}{i}\frac{1}{M^2-p^2+i\Sigma(p^2)-i\epsilon} \\
  G_+^{(1,0)}(p) &= \frac{1}{i}\frac{1}{\omega(\mathbf p)-p^0 +
    i\Sigma_+(p^0,\mathbf p) - i\epsilon}
\end{align}
automatically obey the equation
\begin{equation}
  \Sigma_+(p^0,\mathbf p^2) = \frac{\Sigma_(p^2)}{2\omega(\mathbf p) +
  \frac{i\Sigma(p^2)}{\omega(\mathbf p)+p^0}}
\end{equation}
that was {\em imposed} as a matching condition in the general discussion of
the of two-point functions of a relativistic theory and a non-relativistic
effective theory in appendix~\ref{app:2pf}. Based on this matching, we have
discussed in section~\ref{sec:greens-functions} how off-shell truncated
Green's functions can be matched. The statements made there are true in this
model and we conclude that if we truncate external lines through the function
\begin{equation}
  \label{eq:p-a-new-G-def}
  \mathcal G_+(p) \doteq 
  \frac{G_+^{(1,0)}(p)}{\sqrt{2\omega}}
  \left(1-\frac{i\Sigma_+(p^0,\mathbf p^2)}{\omega(\mathbf p) + p^0}\right)
\end{equation}
according to
\begin{equation}
  \label{eq:p-a-truncation-rule}
  G_+^{(a,b)}(p,q,k) = \hat \mathcal G_+(p)
  \hat \mathcal G_+(q) \hat G_+^{(0,2)}(k) \bar G_{+,tr}^{(a,b)}(p,q,k),
\end{equation}
the equation
\begin{equation}
  \label{eq:p-a-trunc-relation}
  G_{tr}^{(a,b)}(p,q,k) = \bar G_{+,tr}^{(a,b)}(p,q,k)
\end{equation}
is true to all orders in perturbation theory. Furthermore, the residues
$Z_\hf$ and $Z_+$ of $G^{(1,0)}$ and $G_+^{(1,0)}$ are related by
\begin{equation}
  \label{eq:p-a-residues}
  Z_+(\mathbf p^2) = \frac{(\omega(\mathbf p)
  + \op(\mathbf p))^2}{4\op(\mathbf p)\omega(\mathbf p)} Z_\hf.
\end{equation}

\subsection{Amplitudes}
%----------------------
As a consequence of eqns.~\eqref{eq:p-a-trunc-relation}
and~\eqref{eq:p-a-residues}, the on-shell relation
\begin{multline}
  \label{eq:on-shell-equiv-particle}
  Z_\hf^a Z_\lf^{\frac{b}{2}}
  \left. G_{tr}^{(a,b)}(p,q,k)\right|_{\text{on-shell}} = \\
  \prod_{i=1}^{a}\left(Z_+(\mathbf p_i^2) 2\op(\mathbf
    p_i)\right)^{\frac{1}{2}} \left(Z_+(\mathbf q_i^2) 2\op(\mathbf
    q_i)\right)^{\frac{1}{2}} Z_\lf^{\frac{b}{2}} \left. G_{+,tr}^{(a,b)}
    (p,q,k)\right|_{\text{on-shell}},
\end{multline}
where $p_i^0=\omega_{\text{p}}(\mathbf p_i), q_i^0=\omega_{\text{p}}(\mathbf
q_i)$ and $k_i^0=\sqrt{m_{\text{p}}^2+\mathbf k^2}$ is also true. According to
the LSZ formalism, the l.h.s. is related to the amplitude of the process where
$a$ heavy particles scatter into $a$ heavy and $b$ light
particles\footnote{Due to the convention of the Fourier transform given in
  appendix~\ref{app:notation} the momenta $k_i$ with
  $k_i^0=\sqrt{m_{\text{p}}^2+\mathbf k^2}$ correspond to outgoing light
  particles. The amplitude for processes with incoming light particles can be
  obtained by crossing}
\begin{multline}
  \sprod{p_1,\dots,p_a,k_1,\dots,k_b;\text{out}}{q_1,\dots,q_a;\text{in}}
  = \\
  \sprod{p_1,\dots,p_a,k_1,\dots,k_b;\text{in}}{q_1,\dots,q_a;\text{in}}\\
  + i(2\pi)^4\delta^4\left(P+K-Q\right) T_{a\rightarrow a+b},
\end{multline}
where $P=\sum_{i=1}^a p_i$ etc.\ ,through
\begin{equation}
  \label{eq:Taab-rel}
  T_{a\rightarrow a+b} = \frac{1}{i}Z_\hf^a Z_\lf^{\frac{b}{2}}
  \left. G_{tr}^{(a,b)}(p,q,k)\right|_{\text{on-shell}}.
\end{equation}
The same amplitude in the effective theory is given by
\begin{equation}
  \label{eq:Taab-eff}
  T_{a\rightarrow a+b}^+ = \frac{1}{i}\prod_{i=1}^aZ_+(\mathbf
  p_i)^\frac{1}{2} Z_+(\mathbf q_i)^\frac{1}{2} Z_\lf^{\frac{b}{2}}
  \left. G_{+,tr}^{(a,b)}(p,q,k)\right|_{\text{on-shell}}
\end{equation}
and eq.~\ref{eq:on-shell-equiv-particle} is simply the statement that
\begin{equation}
  T_{a\rightarrow a+b} = \prod_{i=1}^a\sqrt{2\op(\mathbf p_i)}
    \sqrt{2\op(\mathbf q_i)} T_{a\rightarrow a+b}^+,
\end{equation}
which is nothing but the matching condition stated in
section~\ref{sec:amplitude-matching}.

We can repeat this procedure with the Lagrangian
\begin{equation}
  \label{eq:L-hm}
  \Lam = \hfm^*\mathcal D_-\hfm + \La_\lf^0,
\end{equation}
describing the anti-particle sector of the theory. In the relativistic theory,
the amplitude for the process where all particles are replaced by
anti-particles is obtained by a simple change of sign of the momenta $p$ and
$q$ as a consequence of crossing symmetry. In the effective theory, however,
the crossed process is described by its own amplitude $G_{-,tr}^{(a,b)}$ and
we get (on-shell has the same meaning as above)
\begin{multline}
  \label{eq:on-shell-equiv-anti-particle}
  Z_\hf^a Z_\lf^{\frac{b}{2}}
  \left. G_{tr}^{(a,b)}(-p,-q,k)\right|_{\text{on-shell}} = \\
  \prod_{i=1}^{a}\left(Z_-(\mathbf p_i^2) 2\op(\mathbf
    p_i)\right)^{\frac{1}{2}} \left(Z_-(\mathbf q_i^2) 2\op(\mathbf
    q_i)\right)^{\frac{1}{2}} Z_\lf^{\frac{b}{2}} \left. G_{-,tr}^{(a,b)}
    (-p,-q,k)\right|_{\text{on-shell}}.
\end{multline}
The connection with the amplitudes $T_{\bar a\rightarrow\bar a+b}$,$T_{\bar
  a\rightarrow\bar a+b}^+$ of the scattering of $a$ anti-particles into $a$
anti-particles and $b$ light particles is analogous to
eqns.~\eqref{eq:Taab-rel} and~\eqref{eq:Taab-eff} and we arrive at the same
conclusions as above.

We have demonstrated in this section that the non-local Lagrangians $\La_\pm$
defined in eqns.~\eqref{eq:L-hp} and~\eqref{eq:L-hm} generate scattering
amplitudes in the pure particle- and anti-particle sector (including any
number of light particles) that are related to the corresponding quantities in
the full theory by the matching condition described in
section~\ref{sec:amplitude-matching}. Strictly speaking, the expressions given
in eqns.~\eqref{eq:on-shell-equiv-particle}
and~\eqref{eq:on-shell-equiv-anti-particle} are valid to all orders in
perturbation theory only in the presence of a regulator that renders all loops
finite. However, the non-local theory is related so closely to the original
one that it is evident that once the full theory is renormalized to some order
in $e$, these expressions are valid up to the same order, because the very
same counter terms render both theories finite at the same time (see
appendix~\ref{app:1-loop-renormalization} for the explicit renormalization to
one loop).

\subsection{Comment on the Structure of Green's Functions}
\begin{figure}[thbp]
  \begin{center}
    \leavevmode
    \includegraphics[width=\textwidth]{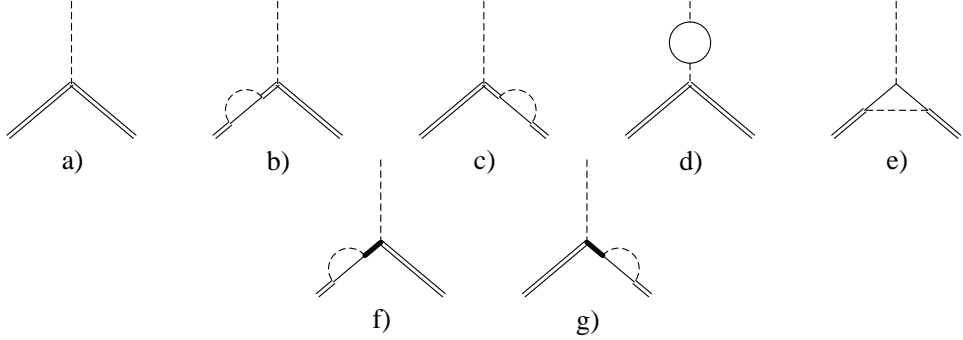}
    \caption{The graphs contributing to the 3-point function $G_+^{(1,1)}$ to
      $O(e^3)$. The solid and dashed lines represent propagators $\Delta_M$
      and $\Delta_m$, the double line particle propagators
      $1/(2\omega)\Delta_+$ and the thick solid line anti-particle propagators
      $1/(2\omega)\Delta_-$. External heavy lines are multiplied with an
      additional factor of $\sqrt{2\omega}$ so that they effectively
      correspond to $1/\sqrt{2\omega}\Delta_+$.}
    \label{fig:G3-eff}
  \end{center}
\end{figure}

%---------------------------------------------------------
The seemingly complicated relation ~\eqref{eq:p-a-trunc-relation} between the
Green's functions of the relativistic and the effective theory is in fact
quite simple. Let us illustrate this with the 3-point functions $G^{(1,1)}$
and $G_+^{(1,1)}$ to $O(e^3)$. The former can be depicted as the sum of the
graphs\footnote{We omit all tadpole graphs in accordance with the
  1-loop renormalization discussed in
  appendix~\ref{app:1-loop-renormalization}} of figure~\ref{fig:G3-rel}.
\begin{figure}[thbp]
  \begin{center}
    \leavevmode
    \includegraphics[width=\textwidth]{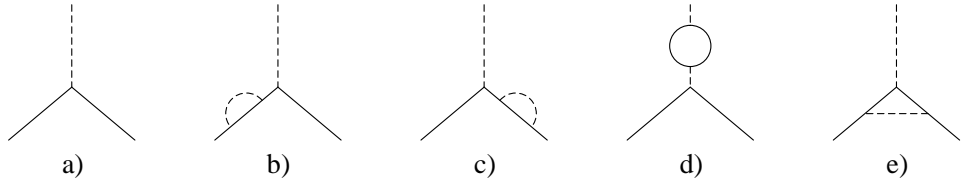}
    \caption{The graphs contributing to the 3-point function $G^{(1,1)}$ to
      $O(e^3)$. The solid and dashed lines represent propagators $\Delta_M$
      and $\Delta_m$, respectively.}
    \label{fig:G3-rel}
  \end{center}
\end{figure}
The corresponding function $G_+^{(1,1)}$ can be obtained from these graphs by
the following simple rules.
\begin{itemize}
\item Replace all internal propagators $\Delta_M(p)$ by the sum
  \begin{displaymath}
    \Delta_M(p) = -\frac{1}{2\omega(\mathbf p)}\left( \Delta_+(p) +
    \Delta_-(p) \right).
  \end{displaymath}
\item Replace all external heavy propagators by particle propagators according
  to
  \begin{displaymath}
    \Delta_M(p) \rightarrow \frac{1}{\sqrt{2\omega(\mathbf p)}} \Delta_+(p).
  \end{displaymath}
\end{itemize}
The resulting graphs are shown in figure~\ref{fig:G3-eff}. It is convenient to
display the decomposition of $\Delta_M$ only for the lines that connect
1-particle irreducible subgraphs.
The meaning of the truncation rule in eq.~\eqref{eq:p-a-truncation-rule}
becomes now apparent. The function $G_{tr}^{(1,1)}$ is given by the sum of
graphs a) and e) of figure~\ref{fig:G3-rel} with external lines removed. The
``naturally'' truncated function $G_{+,tr}^{(1,1)}$, however, is the sum of
graphs a), e), f) and g) with external factors of $\Delta_+$ and $\Delta_m$
removed. The point is that some parts that belong to insertions on the heavy
external lines in the relativistic theory are now considered to belong to the
irreducible vertex function because the anti-particle propagator $\Delta_-$ is
considered to be irreducible. The modified truncation rule, involving
$\mathcal G$ defined in eq.~\eqref{eq:p-a-new-G-def}, on the other hand gives
the truncated function $\bar G_{+,tr}^{(1,1)}$ which only contains graphs a)
and e).  We have thus verified explicitly the equation
\begin{displaymath}
  G_{tr}^{(1,1)} = \bar G_{+,tr}^{(1,1)}
\end{displaymath}
to $O(e^3)$.

%%% Local Variables: 
%%% mode: latex
%%% TeX-master: "~/diss/tex/diss"
%%% End: 
\chapter{$1/M$ Expansion}
%=======================
\markboth{$1/M$ EXPANSION}{}
\label{chap:M-expansion}
The Lagrangians constructed in the preceding chapter are non-local, i.e. they
depend on the entire configuration space. The explicit expression for
$\mathcal L_+$ is
\begin{multline}
  \label{eq:L-hp-non-local}
  \mathcal L_+(x) = \int
  d^4y\hfp^*(x)\left(\delta^4(x-y)(D_{+,y}-eB(y))- \right. \\
  \left. e^2B(x)C^{-1}(x,y)B(y)\hfp(y)\right).
\end{multline}
We showed that this theory contains the same truncated Green's functions as
the original local field theory. The whole purpose of the construction of
$\mathcal L_\pm$ is to pave the way for the expansion of these Green's
functions in the region where all energies and momenta are small compared to
the mass $M$. This expansion turns the non-local Lagrangians into local ones,
which should be able to reproduce the expansion of relativistic Green's
functions. 

In this chapter, we first look at a few simple processes in the
relativistic theory and discuss their $1/M$ expansion at tree level. Then we
perform the expansion in the non-local Lagrangian and discuss how perturbation
theory works. Finally, we check the method in the case of the scattering of a
heavy and a light particle at tree level.

\section{Expansion of Relativistic Amplitudes at Tree Level}
%--------------------------------------------------------------
\label{sec:on-shell-expansion}
We consider the truncated Green's functions $G_{tr}^{(2,0)}$ and
$G_{tr}^{(1,2)}$ on the mass shell, i.e. the heavy momenta obey $p^2=\Mp^2$
and the light momenta $k^2=m_{\text{p}}^2$.

\subsection{Heavy-Heavy Scattering}
\begin{figure}[thbp]
  \begin{center}
    \leavevmode
    \includegraphics[width=\textwidth]{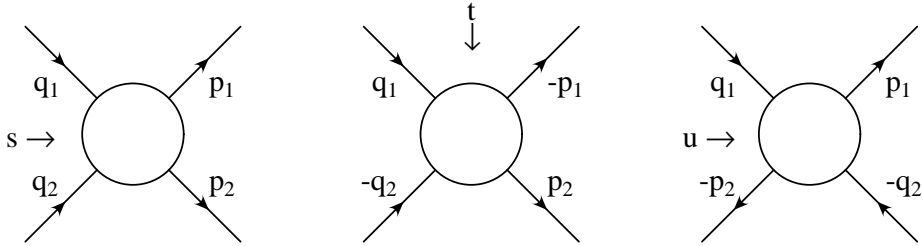}
    \caption{Physical processes associated with the amplitude $A(s,t,u)$
      defined in eq.~\eqref{eq:heavy-heavy-amplitude}. In the $s$-channel it
      describes the scattering of two heavy particles and in the $t$- and
      $u$-channels the scattering of a particle and an anti-particle. The
      lines are labeled by the physical momenta in the respective channels.}
    \label{fig:heavy-heavy-scattering}
  \end{center}
\end{figure}

\begin{figure}[thbp]
  \begin{center}
    \leavevmode \includegraphics{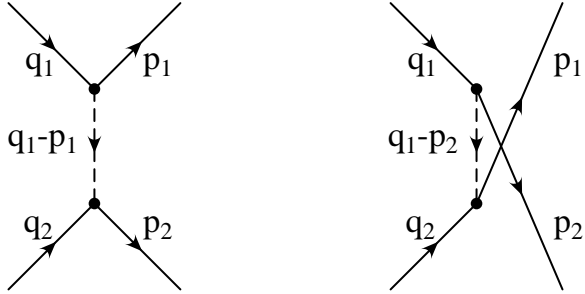}
    \caption{The graphs that contribute to $A(s,t,u)$ at tree-level.}
    \label{fig:heavy-heavy-tree}
  \end{center}
\end{figure}

%----------------------------------
The function $G_{tr}^{(2,0)}(p_1,p_2,q_1,q_2)$ involves only heavy external
particles. With the convention for the Fourier transform of Green's functions
given in appendix~\ref{app:notation}, $q_1$,$q_2$ are the physical momenta of
incoming particles and $p_1$,$p_2$ those of outgoing ones. Therefore,
$q_1+q_2$ is the total energy in the CMS of particle-particle scattering. We
define the Mandelstam variables
\begin{align}
  s &= (q_1 + q_2)^2 \notag \\
  t &= (q_1 - p_1)^2 \notag \\
  u &= (q_1 - p_2)^2
\end{align}
related by
\begin{equation}
  s+t+u = 4\Mp^2.
\end{equation}
The invariant amplitude
\begin{equation}
  \label{eq:heavy-heavy-amplitude}
  A(s,t,u) = \frac{1}{i} Z_\hf^2
  \left. G_{tr}^{(2,0)}(p_1,p_2,q_1,q_2)\right|_{\text{on-shell}}  
\end{equation}
describes several physical processes in different regions of momentum space
(cf. figure~\ref{fig:heavy-heavy-scattering}). We define the amplitudes
belonging to the various channels by
\begin{align}
  \label{eq:heavy-heavy-channels}
  A_s(s,t,u) &= \left.A(s,t,u)\right|_{q_1^0,q_2^0,p_1^0,p_2^0>0} \notag \\
  A_t(t,s,u) &= \left.A(s,t,u)\right|_{q_1^0,p_2^0>0;q_2^0,p_1^0<0} \notag \\
  A_u(u,t,s) &= \left.A(s,t,u)\right|_{q_1^0,p_1^0>0;q_2^0,p_2^0<0},
\end{align}
writing the energy in the CMS and the momentum transfer as the first and
second arguments, respectively.  In the $s$-channel, $A(s,t,u)$ describes
particle-particle scattering and in the $t$- and $u$-channels
particle-anti-particle scattering.
The presence of identical particles is reflected in the crossing symmetry
\begin{equation}
  A(s,t,u) = A(s,u,t).
\end{equation}
In perturbation theory we write
\begin{equation}
  A(s,t,u) = e^2 A^{(2)}(s,t,u) + O(e^4)
\end{equation}
and find that the lowest order is given by the two tree-level Feynman diagrams
shown in figure~\ref{fig:heavy-heavy-tree}
\begin{equation}
  \label{eq:A2}
  A^{(2)}(s,t,u) = \frac{1}{m^2-t} + \frac{1}{m^2-u}.
\end{equation}
Let us expand this quantity for the case when all three-momenta as well as $m$
are much smaller than the heavy scale $M$. This expansion has to be performed
separately in each channel and we start with the $s$-channel. It is convenient
to work in the CMS, where $q_1=(\sqrt{s}/2,\mathbf q)$,
$q_2=(\sqrt{s}/2,-\mathbf q)$, $p_1=(\sqrt{s}/2,\mathbf p)$ and
$p_2=(\sqrt{s}/2,-\mathbf p)$. The CM energy $s$ is of the order of $4M^2$ and
thus represents a hard scale, where as the momentum transfer $t$ and $u = 4M^2
- s-t$ are soft
\begin{align}
  t &= -(\mathbf q - \mathbf p)^2 \\
  u &= -(\mathbf q + \mathbf p)^2
\end{align}
Therefore, both denominators in eq.~\eqref{eq:A2} are small and 
\begin{equation}
  A_s^{(2)}(s,t,u) = \frac{1}{m^2+(\mathbf q - \mathbf p)^2} + 
  \frac{1}{m^2+(\mathbf q + \mathbf p)^2}.
\end{equation}

In the $t$-channel, $t$ is the hard CM energy. In the CMS, where
$q_1=(\sqrt{t}/2,\mathbf q)$, $p_1=(-\sqrt{t}/2,\mathbf q)$,
$q_2=(-\sqrt{t}/2,\mathbf p)$ and $p_2=(\sqrt{t}/2,\mathbf p)$, we have
\begin{equation}
  t = 4(M^2 + \mathbf q^2).
\end{equation}
The momentum transfer $s$ and
\begin{equation}
  u = -(\mathbf q - \mathbf p)^2
\end{equation}
are still soft. In this channel, the first graph of
figure~\ref{fig:heavy-heavy-tree} represents an annihilation process, where
the particle and anti-particle convert into a light particle which is then
considerably off its mass shell, followed by pair production. The leading term
of the expanded propagator is of $O(1/M^2)$ and indicates that this process
looks essentially point-like in configuration space on a scale much larger
than $1/M$. The second term involves only the exchange of soft momenta and has
a leading piece that is not suppressed by powers of $1/M$
\begin{equation}
  \label{eq:At2}
  A_t^{(2)}(t,s,u) = \frac{1}{m^2+(\mathbf q - \mathbf p)^2} -
  \frac{1}{4M^2}\left(1-\frac{4\mathbf q^2-m^2}{4M^2} +
  O(\frac{1}{M^4})\right). 
\end{equation}

\subsection{Heavy-Light Scattering}
\begin{figure}[thbp]
  \begin{center}
    \leavevmode \includegraphics{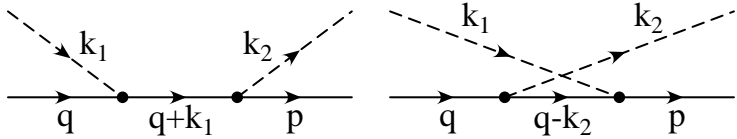}
    \caption{The graphs that contribute to $B(s,t,u)$ at tree-level.}
    \label{fig:heavy-light-tree}
  \end{center}
\end{figure}

\begin{figure}[thbp]
  \begin{center}
    \leavevmode \includegraphics{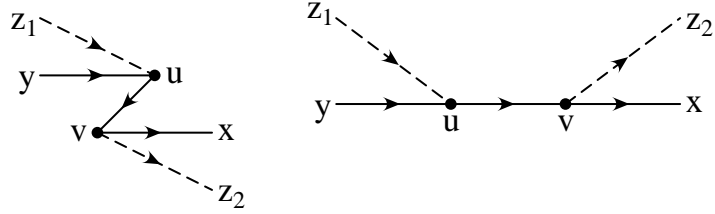}
    \caption{Decomposition of the first graph of
      figure~\ref{fig:heavy-light-tree} according to
      eq.~\eqref{eq:prop-decomposition} in configuration space (an integration
      over the internal points $u$,$v$ is implied).}
    \label{fig:heavy-light-z-graph}
  \end{center}
\end{figure}

%-------------------------------------
Let us chose the momentum assignment in the Fourier transform of $G^{(1,2)}$
as follows
\begin{multline}
  (2\pi)^4\delta^4(p+k_2-q-k_1)G^{(1,2)}(p,q,k_1,k_2) = \\
  \int d^4xd^4yd^4z_1d^4z_2 e^{ipx - iqy + ik_2z_2 -
    ik_1z_1}G^{(1,2)}(x,y,z_1,z_2).
\end{multline}
With this choice, $q$,$k_1$ are the physical momenta of incoming particles and
$p$,$k_2$ those of outgoing ones. Therefore, $q+k_1$ is the total energy in
the CMS of the process where a light particle scatters off a heavy one and we
chose the Mandelstam variables
\begin{align}
  s &= (q + k_1)^2 \notag \\
  t &= (q - p)^2 \notag \\
  u &= (q - k_2)^2
\end{align}
with
\begin{equation}
  s+t+u = 2(\Mp^2 + m_{\text{p}}^2).
\end{equation}
The different processes represented by
\begin{equation}
  \label{eq:heavy-light-amplitude}
  B(s,t,u) = \frac{1}{i} Z_\hf Z_\lf
  \left. G_{tr}^{(1,2)}(p,q,k_1,k_2)\right|_{\text{on-shell}}  
\end{equation}
\begin{figure}[thbp]
  \begin{center}
    \leavevmode
    \includegraphics[width=\textwidth]{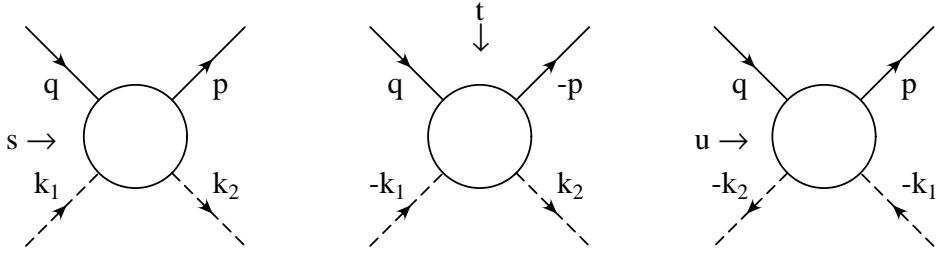}
    \caption{Physical processes associated with the amplitude $B(s,t,u)$
      defined in eq.~\eqref{eq:heavy-light-amplitude}. Solid and dashed lines
      represent heavy and light particles, respectively. In the $s$- and $u$-
      channels it describes scattering and in the $t$-channel
      pair-annihilation.  The lines are labeled by the physical momenta in the
      respective channels.}
    \label{fig:heavy-light-scattering}
  \end{center}
\end{figure}
are shown in figure~\ref{fig:heavy-light-scattering} and the amplitudes in the
different channels are defined in analogy with
eq.~\eqref{eq:heavy-heavy-channels}
\begin{align}
  \label{eq:heavy-light-channels}
  B_s(s,t,u) &= \left.B(s,t,u)\right|_{q^0,p^0,k_1^0,k_2^0>0} \notag \\
  B_t(t,s,u) &= \left.B(s,t,u)\right|_{q^0,k_2^0>0;p^0,k_1^0<0} \notag \\
  B_u(u,t,s) &= \left.B(s,t,u)\right|_{q^0,p^0>0;k_1^0,k_2^0<0}.
\end{align}
We refer to the $s$- and $u$-channels as Compton scattering and the
$t$-channel as pair-annihilation. Because of the crossing symmetry
\begin{equation}
  B(s,t,u) = B(u,t,s),
\end{equation}
we can again restrict the analysis to the $s$- and $t$-channels. Let us set
\begin{equation}
  B(s,t,u) = e^2 B^{(2)}(s,t,u) + O(e^4)
\end{equation}
where $B^{(2)}$ is given by the Feynman diagrams displayed in
figure~\ref{fig:heavy-light-tree}
\begin{equation}
  B^{(2)}(s,t,u) = \frac{1}{M^2-s} + \frac{1}{M^2-u}.
\end{equation}
In contrast to the processes considered above, this amplitude explicitly
depends on the heavy scale through the propagator
\begin{equation}
  \Delta_M(p) = \frac{1}{M^2-p^2}.
\end{equation}
The construction of the non-local Lagrangians $\La_\pm$ relied essentially on
the decomposition
\begin{equation}
  \label{eq:prop-decomposition}
  \Delta_M(p) = \frac{1}{2\omega(\mathbf p)}\left( \frac{1}{\omega(\mathbf
  p)-p^0} + \frac{1}{\omega(\mathbf p)+p^0} \right).
\end{equation}
of this function, representing the propagation of a particle and an
anti-particle separately.  The important point is that when $p^0$ is in the
vicinity of $+\omega(\mathbf p)$, the first term dominates where as the second
one can be expanded in powers of $1/M$ and vice versa if $p^0$ is in the
vicinity of $-\omega(\mathbf p)$. In configuration space, the first graph of
figure~\ref{fig:heavy-light-tree} may be depicted as the sum of the two graphs
in figure~\ref{fig:heavy-light-z-graph}.
In these diagrams, the internal propagators correspond to factors
$d^2\Delta_-(v-u)$ and $d^2\Delta_+(u-v)$, respectively (cf.
appendix~\ref{app:KG} for the definition of these objects).  Due to its shape,
the first graph is called a ``Z'' graph in the language of old-fashioned
(non-covariant) perturbation theory. 

In the $s$-channel, the incident light
particle pushes the incident heavy particle only slightly off the mass shell,
so that the internal anti-particle propagator in the $Z$ graph is far away
from its pole at $p^0=-\omega(\mathbf p)$ and is suppressed relative to the
other graph. The $Z$ graph looks like an effective local four-particle
interaction
\begin{displaymath}
  \includegraphics{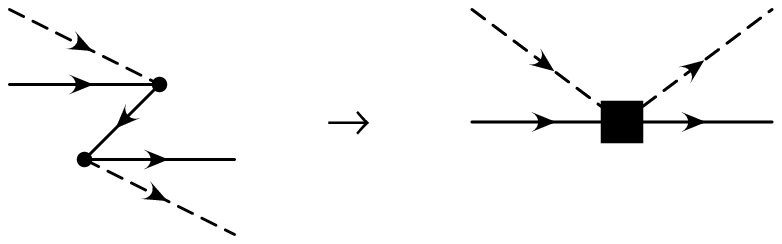}
\end{displaymath}
Let us work in the rest frame of the incoming heavy particle where $q=(M,0)$,
$k_1=(\Omega(\mathbf k_1),\mathbf k_1)$, $p=(\omega(\mathbf p),\mathbf p)$,
$k_2=(\Omega(\mathbf k_2),\mathbf k_2)$ and $\Omega(\mathbf
k)=\sqrt{m^2+\mathbf k^2}$. The contribution to the amplitude
$B_s^{(2)}(s,t,u)$ of the Z graph is
\begin{multline}
  \frac{1}{2\omega(\mathbf k_1)} \frac{1}{\omega(\mathbf k_1) +
  \Omega(\mathbf 
  k_1) + M} = 
  \frac{1}{4M^2}\left( 1 - \frac{\Omega(\mathbf k_1)}{2M} + O(\frac{1}{M^2})
  \right). 
\end{multline}
The other part of the diagram gives the leading contribution
\begin{multline}
  \frac{1}{2\omega(\mathbf k_1)} \frac{1}{\omega(\mathbf k_1) -
  \Omega(\mathbf k_1) - M} = \\ 
  \frac{-1}{2M\Omega(\mathbf k_1)}\left( 1 + \frac{\mathbf
  k_1^2}{2M\Omega(\mathbf k_1)} + O(\frac{1}{M^2})\right)
\end{multline}
and
\begin{align}
  \label{eq:Bs2}
  B_s^{(2)}(s,t,u) =&\; \frac{-1}{2M\Omega(\mathbf k_1)}\left( 1 +
  \frac{\mathbf k_1^2}{2M\Omega(\mathbf k_1)}\right) 
+ \frac{1}{4M^2}\left( 1 - \frac{\Omega(\mathbf k_1)}{2M}\right) \notag
  \\ 
  &\; + (\mathbf k_1
  \rightarrow -\mathbf k_2,\Omega(\mathbf k_1) \rightarrow -\Omega(\mathbf
  k_2) ) + O(\frac{1}{M^2}).
\end{align}

In the $t$-channel, things are different again. Let us chose the CMS and go to
the threshold, where $q=(M,0)$, $p=(-M,0)$, $k_1=(-M,\mathbf k)$ and
$k_2=(M,\mathbf k)$ (remember that the physical momenta are $q$, $-p$, $-k_1$
and $k_2$). The invariants have the values $s=u=m^2-M^2$ and
\begin{equation}
  \label{eq:annihilation-expansion}
  B_t(u,t,s) = \frac{2}{2M^2-m^2} = \frac{1}{M^2}\left(1+\frac{m^2}{2M^2} +
  O(\frac{1}{M^4}) \right).
\end{equation}
This means that pair-annihilation has no soft component: the entire process
looks local on a scale much larger than $1/M$.

To summarize, we may group all processes we have just discussed into three
categories.  If the initial and final states contain exclusively either heavy
particles or anti-particles, we call it a {\em soft} process (charge
conservation implies that the number of particles is conserved). If the
initial and final states contain both types of particles but their number is
separately conserved, we call it a {\em semi-hard} process. Finally, if the
numbers of particles and anti-particles are not conserved separately, we call
it a {\em hard} process.  The number of light particles is not important for
this classification.

\begin{itemize}
\item Soft processes. This category comprises the $s$-channel of the amplitude
  $A$ and the $s$- and $u$-channels of the amplitude $B$ (i.e.
  particle-particle and Compton scattering).  They have in common that at each
  vertex of the tree-level diagrams, only energies and momenta that are much
  smaller than $M$ are transferred. This means that all virtual light
  particles are not far from the mass shell, mediating the interaction over
  distances that are not small compared to $1/M$, and all virtual heavy
  particles are in the vicinity of the particle mass shell, i.e.  the energy
  component of its momentum is close to $\omega$. Therefore, only the
  anti-particle components of these propagators represent a local interaction.
  As a consequence, no more than two heavy lines are attached to a local
  effective vertex.
\item Semi-hard processes. The $t$- and $u$ channels, describing
  particle-anti-particle scattering, of the amplitude $A$ are the only members
  of this category. In the annihilation channel (the first graph of
  figure~\ref{fig:heavy-heavy-tree} in the $t$-channel and the second graph in
  the $u$-channel) a heavy particle annihilates with a heavy anti-particle,
  emitting a virtual light particle that is well off its mass shell and
  travels only a distance of the order of $1/M$, giving rise to local
  interactions with more than two heavy particles involved. The other
  contribution to the process is soft in the sense described above.
\item Hard processes. The pair-annihilation (the $t$-channel of the amplitude
  $B$) is completely local because the virtual particles are always far away
  from the mass shell. This can be traced to the fact that at least one of the
  emerging light particles must be hard: even at threshold, the energy
  released by the annihilating particles is of the order of $M$.
\end{itemize}

Let us discuss the hard processes in more detail.  In the terminology just
established, the process where two heavy particles annihilate into, say, 100
light particles is still considered to be hard. One may think that this is not
adequate, because each of the light particles can be very soft.  However,
there are still some regions of phase space where a sizeable fraction of the
energy is distributed among a few of them, which are then hard. Thus, the
expansion of internal heavy lines depends on the configuration of the final
states and it seems that there is no expansion that is valid everywhere in
phase space. One may say that some pieces of the amplitude require one to
treat both, particle and anti-particle as heavy degrees of freedom. The point
is that neither $\Lap$ nor $\Lam$ are valid in this region. 

Looking at equation~\eqref{eq:annihilation-expansion}, one might be tempted to
simply add a local interaction of the type $\hfp\hfm^*\lf^2$ (and its
hermitian conjugate), since the entire process is local. Such a term
contributes also to the two-point function $\eval{0}{T\hfp(x)\hfp^\dag(y)}{0}$
at $O(e^4)$. Now, this Green's function is already correctly described by
$\Lap$ alone, as we have seen in section~\ref{sec:p-a-sector}, and there
arises the problem of double counting: by adding the mentioned local term, we
must change the coefficients of $\Lap$ already fixed by a matching in the
particle sector. It is a priori not clear if this procedure can be implemented
systematically.

In addition, unitarity tells us that the
tree-level amplitudes of figure~\ref{fig:heavy-light-tree} in the $t$ channel
are related to the imaginary part of the diagram
\begin{displaymath}
  \includegraphics{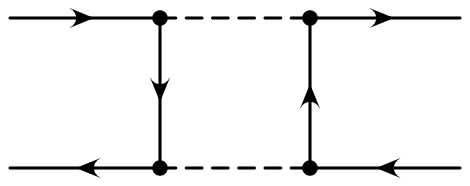}
\end{displaymath}
in the particle-anti-particle channel. Being a semi-hard process, we
expect that the box is represented as a string of local four-particle (two
particle and two anti-particle) interactions. This is again in conflict with a
term of the form $\hfp\hfm^*\lf^2$, because two of these vertices essentially
generate the box itself. 

These are the reasons why annihilation processes are usually excluded from the
effective Lagrangian. Attempts have been made to include them in order to
describe positronium decay~\cite{labelle-thesis} or heavy quarkonium
decay~\cite{hqqbaret}.

Clearly, this subject deserves further investigation.

\section{Off-Shell Expansion}
%-------------------------------
\label{sec:off-shell-expansion}
In section~\ref{sec:p-a-sector} we have seen that we can reproduce the
truncated off-shell Green's functions of the relativistic theory if we use a
special truncation prescription in the effective theory, which amounts to
multiply with an additional factor $\sqrt{2\omega}$ for every external heavy
line at tree level. The local effective theory is expected to produce a $1/M$
expansion of Green's functions. Let us therefore extend the expansion
discussed in the last section to off-shell momenta, i.e. we go back to the
functions $G_{tr}^{(2,0)}$ and $G_{tr}^{(1,2)}$, treating the energy
components of the momenta as independent variables.

\subsection{Heavy-Heavy scattering}
%-------------------------------------
We keep the notation with Mandelstam variables but discard the on-shell
conditions. Strictly speaking, we cannot talk about different channels any
more because we are outside of the physical region. However, to stay in the
scope of a $1/M$ expansion, we cannot move too far away from the mass shell so
that the notion of channels still has some meaning. In the $s$-channel, for
example, we restrict the energies of the particles to be much smaller than $M$
in the sense that $|q_i^0-M|$, $|p_1^0-M| \ll M$. It is convenient to
introduce new variables ($i=1,2$)
\begin{align}
  E_{q_i} &\doteq q_i^0-M \\ 
  E_{p_i} &\doteq p_i^0-M. 
\end{align}
The Green's function depends on several small dimensionless quantities
$E_{q_i}/M$, $|\mathbf q_i|/M$, \dots and we must decide what their relative
magnitude is. At the moment, we do not have any preference and simply consider
all of them to be of equal magnitude, which is the same as counting powers of
$1/M$ as before. More about this issue will be said below. In this framework,
the function
\begin{multline}
  \frac{1}{i} G_{tr}^{(2,0)}(p_1,p_2,q_1,q_2) = \frac{1}{m^2 -
  (E_{q_1}-E_{p_1})^2 + (\mathbf q_1 - \mathbf p_1)^2} \\
+ \frac{1}{m^2 - (E_{q_1}-E_{p_2})^2 + (\mathbf q_1 - \mathbf p_2)^2}
\end{multline}
cannot be expanded at all. 

In the $t$-channel, $|p_1^0-M|$ and $|q_2^0-M|$ are of the order of $M$. The
good variables are in this case
\begin{align}
  \bar E_{p_1} &\doteq p_1^0+M \\
  \bar E_{q_2} &\doteq q_2^0+M
\end{align}
in the sense that $|\bar E_{p_1}|$, $|\bar E_{q_2}|\ll M$. We find
\begin{multline}
  \label{eq:h-h-t-off-shell}
  \frac{1}{i}G_{tr}^{(2,0)}(p_1,p_2,q_1,q_2) = \frac{1}{m^2 -
  (E_{q_1}-E_{p_2})^2 + (\mathbf q_1 - \mathbf p_2)^2} \\
- \frac{1}{4M^2}\left(1 - \frac{E_{q_1}-\bar
  E_{p_1}}{M} + \frac{3(E_{q_1}-\bar E_{p_1})^2 + (\mathbf q_1 - \mathbf
  p_1)^2 + m^2}{4M^2} + O(\frac{1}{M^3}) \right).
\end{multline}

\subsection{Heavy-Light scattering}
%-------------------------------------
Using the same energy variables as before and considering the energy
components of the light momenta to be of the same order, we find in the
$s$-channel
\begin{multline}
  \label{eq:Gtr12-HM-expanded}
  \frac{1}{i} G_{tr}^{(1,2)}(p,q,k_1,k_2) = \frac{-1}{2M(E_q+k_1^0)}\left(
  1 + \frac{(\mathbf q + \mathbf k_1)^2}{2M(E_q+k_1^0)} + \right. \\
  \left. + \frac{(\mathbf q + \mathbf k_1)^4}{4M^2(E_q+k_1^0)^2}
  - \frac{(\mathbf q + \mathbf k_1)^2}{4M^2} 
  O(\frac{1}{M^3}) \right) \\
+ \frac{1}{4M^2}\left(1-\frac{E_q+k_1^0}{2M} + O(\frac{1}{M^2}) \right) +
  (k_1\rightarrow - k_2).
\end{multline}

\section{Effective Local Lagrangians for Soft Processes}
%----------------------------------------------------------
\label{sec:M-expansion-of-L}
The non-local theories constructed in section~\ref{sec:p-a-sector} are
naturally restricted to soft processes in the particle and anti-particle
sectors and we have proven that they reproduce the relativistic theory exactly
at tree level. The effective local Lagrangians are obtained by expanding the
non-local pieces, which are the anti-particle propagator $\Delta_-$ and the
particle propagator $\Delta_+$ for $\Lap$ and $\Lam$, respectively. Let us
first concentrate on $\Lap$. We find
\begin{align}
  \Delta_-(x) =\; & -\frac{1}{2M}\left(1-\frac{i\partial_t-M}{2M}
  + \frac{\Delta}{4M^2} \right. \notag \\
  & \left. + \frac{(i\partial_t-M)^2}{4M^2} + O(\frac{1}{M^3})
  \right) \delta^4(x)
\end{align}
and, expanding the operator $d=(2\sqrt{M^2-\Delta})^{-1/2}$ as well, we can
write the Lagrangian in the form
\begin{equation}
  \Lap = \hfp^*D_+\hfp + \sum_{n=1}^\infty \frac{1}{(2M)^n}\Lap^{(n)},
\end{equation}
where
\begin{align}
  \Lap^{(1)} =&\; e\hfp^*\lf\hfp \\
  \Lap^{(2)} =&\; 0 \\
  \Lap^{(3)} =&\; \hfp^*\left\{ e(\lf\Delta + \Delta\lf) + e^2\lf^2
  \right\}\hfp \\ 
  \Lap^{(4)} =&\; -e^2\hfp^*\lf(i\partial_t-M)\lf\hfp \\
  \Lap^{(5)} =&\; \hfp^*\left\{ e\left(\Delta\lf\Delta +
  \frac{5}{2}\lf\Delta^2 + 
  \frac{5}{2}\Delta^2\lf\right) \right. \notag \\
& \left. + e^2\left(\lf^2\Delta + \Delta\lf^2 + 3\lf\Delta\lf 
    + \lf[i\partial_t - M]^2\lf
  \right)\right. \notag \\
& \biggr. + e^3\lf^3 \biggr\}\hfp
\end{align}
and the differential operators act on everything on their right. In the
anti-particle sector, the Lagrangian is of the same form and the $\Lam^{(n)}$
are obtained from the $\Lap^{(n)}$ by replacing $\hfp$ by $\hfm$ and
$i\partial_t-M$ by $-i\partial_t-M$.

\subsection{Including Semi-Hard Processes}
%-----------------------------------------
The semi-hard processes contain virtual pair-annihilation and
creation processes, represented by local effective interactions of several
heavy particles. It is clear that a candidate for the effective theory that
should include these reactions must contain both types of heavy
particles. Consider the Lagrangian
\begin{equation}
  \La = \Lap + \Lam.
\end{equation}
It clearly contains the pure particle- and anti-particle sectors as well as
the soft part of the semi-hard particle-anti-particle processes but not the
hard part of the latter. To include those, we must supplement the Lagrangian
with contact interactions between particles and anti-particles of the form
\begin{equation}
  \La_c = \sum_{n=1}^\infty e^{2n} \La_c^{(n)},
\end{equation}
where $\La_c^{(n)}$ contains $n$ factors of the fields $\hfp$,$\hfp^*$,$\hfm$
and $\hfm^*$. Each of these terms is itself an expansion in $1/M$
\begin{equation}
  \La_c^{(n)} = \sum_{m=0}^\infty \frac{1}{(2M)^{4n+m}} \La_c^{(n,m)}.
\end{equation}
The first two terms of $\La_c^{(1)}$ can be read off from the second term of
eq.~\eqref{eq:h-h-t-off-shell} 
\begin{align}
  \La_c^{(1,0)}(x) =&\; \hfp^*(x)\hfp(x)\hfm^*(x)\hfm(x) \\
  \La_c^{(1,1)}(x) =&\; -2\hfp^*(x)\left(
  [(i\partial_{x^0}-M)\hfp(x)]\hfm^*(x) \right. \notag \\
  & \left. + \hfp(x)[(i\partial_{x^0}-M)\hfm^*(x)]\right)\hfm(x).
\end{align}

%%% Local Variables: 
%%% mode: latex
%%% TeX-master: "~/diss/tex/diss"
%%% End: 
\section{Power Counting Schemes}
%-------------------------------
\label{sec:power-counting}
In the relativistic theory, there is only one expansion parameter: the
coupling constant $e$. The effective theory contains many more small
parameters, namely the energies and momenta of the process of interest, which
are considered to be small compared to $M$. In such a multiple expansion, the
question of ordering arises, i.e. what is the relative magnitude of the
expansion parameters, which determines what terms in the expansion should be
grouped together. We refer to a particular ordering as a {\em power counting
  scheme}. In the following we discuss the two schemes which are of practical
importance.

Because the effective theory should reproduce quantities of the fundamental
theory, the primary expansion parameter is the coupling $e$.  If we go to the
mass shell, the energies of the particles are expressed in terms of their
momenta and the number of independent expansion parameters is reduced.  In
section~\ref{sec:on-shell-expansion}, we have expanded some on-shell amplitudes to a
fixed order in $e$ and some power of $1/M$, i.e. we have collected terms with
the same powers of $e$ and $1/M$. Formally, we may introduce a small number
$v$ as a bookkeeping device and assign powers of it to the expansion
parameters after making them dimensionless by dividing with appropriate powers
of $M$. To the momentum $\mathbf p$ of a heavy or a light particle we assign 
\begin{equation}
  \frac{|\mathbf p|}{M} = O(v).
\end{equation}
The energy $\Omega(\mathbf k) = \sqrt{m^2 + \mathbf k^2}$ of a light particle
is counted as
\begin{equation}
  \frac{\Omega(\mathbf k)}{M} = O(v).
\end{equation}
This implies that, formally, $m/M$ is considered to be of the same order as
$|\mathbf k|/M$. As a consequence of these assignments, $|\mathbf
p|/\Omega(\mathbf k)$ is of order one. In this language, we would say, for
example, that the amplitude $B_s^{(2)}$ in eq.~\eqref{eq:Bs2} is correct up to
terms of $O(v^2)$.

In the off-shell expansion performed in section~\ref{sec:off-shell-expansion},
we have simply counted powers of $1/M$. This is equivalent to setting
\begin{align}
  \frac{|E|}{M} &= O(v) & \frac{|k^0|}{M} &= O(v),
\end{align}
where $E$ is the energy component of the four vector of a heavy particle with
the mass $M$ subtracted and $k^0$ the energy of a light particle. Clearly, the
assignment of $E/M$ is not compatible with the one of $|\mathbf p|/M$ if we go
on-shell, because
\begin{equation}
  E = p^0-M = \frac{\mathbf p^2}{2M} + O(M v^4),
\end{equation}
i.e. $E$ becomes a quantity of $O(v^2)$. However, no harm is done, because we
formally consider $E$ to be larger than it actually is on-shell. This can be
seen, for example, in the amplitude $A_t^{(2)}$. The $1/M$ suppressed
contribution to the on-shell function is given by\footnote{In the CMS}
(eq.~\eqref{eq:At2}) 
\begin{equation}
  -\frac{1}{4M^2}\left(1-\frac{4\mathbf q^2-m^2}{4M^2} +
  O(v^4)\right),
\end{equation}
where as the off shell expansion yields\footnote{In no particular frame of reference} (eq.~\eqref{eq:h-h-t-off-shell})
\begin{multline}
  -\frac{1}{4M^2}\left(1 - \frac{E_{q_1}-\bar
  E_{p_1}}{M} + \frac{3(E_{q_1}-\bar E_{p_1})^2 + (\mathbf q_1 - \mathbf
  p_1)^2 + m^2}{4M^2} + O(v^3) \right).
\end{multline}
On-shell we have $E_{q_1}=\frac{\mathbf q_1^2}{2M}+O(Mv^4)$ and $\bar E_{p_1}
= - \frac{\mathbf p_1^2}{2M} + O(Mv^4)$ and, going to the CMS, both expressions
agree to $O(v^2)$.

Of course, we could just as well have performed the off-shell expansion by
setting
\begin{equation}
  \frac{|E|}{M} = O(v^2).
\end{equation}
In this case, we get
\begin{equation}
  -\frac{1}{4M^2}\left(1 - \frac{4M(E_{q_1}-\bar E_{p_1}) - (\mathbf q_1 -
  \mathbf p_1)^2 - m^2}{4M^2} + O(v^4) \right),
\end{equation}
which also agrees with the previous expressions on-shell and to $O(v^2)$.

From this discussion, we can learn two things
\begin{itemize}
\item The expansion of on-shell amplitudes is naturally associated with an
  expansion in $1/M$.
\item There is no natural choice for the expansion of off-shell amplitudes (or
  Green's functions). Counting $E/M$ the same as $|\mathbf p|/M$ conserves the
  strict $1/M$ expansion but the orders get mixed if we go on-shell (the terms
  $E/M$ will contribute to all higher orders). If we count $E/M$ as $|\mathbf
  p|^2/M^2$, we do not expand simply in $1/M$ but the energies are considered
  to be of the order they actually are on-shell. Also in this case does a term
  $E/M$ contribute to all higher orders if we go on-shell. Different counting
  schemes are possible but not of practical importance.
\end{itemize}

To conclude, we define two power counting schemes ($p$ and $k$ denote the
four-momenta of a heavy and a light particle, respectively):
\begin{enumerate}
\item Heavy-Meson (HM) scheme.  Its defining feature is that the
  three-momentum and the energy variable (with the heavy mass subtracted) are
  considered to be of equal magnitude
  \begin{align}
    \frac{|E|}{M} &= O(v) & \frac{|k^0|}{M} = O(v) \\
    \frac{|\mathbf p|}{M} &= O(v) & \frac{|\mathbf k|}{M} = O(v).
  \end{align}
  The name is an adaptation from HBCHPT~\cite{hbchpt-review}, where this
  counting scheme is used.
\item Non-Relativistic (NR) scheme. In this scheme, the energy of a heavy
  particle is counted like its three-momentum squared
  \begin{align}
    \frac{|E|}{M} &= O(v^2) & \frac{|k^0|}{M} = O(v)\\
    \frac{|\mathbf p|}{M} &= O(v) & \frac{|\mathbf k|}{M} = O(v).
  \end{align}
  The name is derived from the fact that the lowest order effective Lagrangian
  is Galilei-invariant and thus represents a true non-relativistic
  theory.
\end{enumerate}

%%% Local Variables: 
%%% mode: latex
%%% TeX-master: "~/diss/tex/diss"
%%% End: 
\section{Perturbation Theory}
%----------------------------
\label{sec:perturbation-theory}
We have seen that the relativistic Green's functions can be expanded in
different ways. What does this mean for the effective theory? The effective
Lagrangian contains space and time derivatives of the fields. In momentum
space they become three-momentum and energy variables and the question of
ordering arises already on the level of the Lagrangian. It is clear that the
perturbation theory looks different for the different counting schemes. The
resulting Green's functions can then be identified with the different
expansions of the relativistic Green's functions.

In section~\ref{sec:M-expansion-of-L}, we have ordered the effective Lagrangian
according to powers of $1/M$. It is useful to reorder it now. First of all, we
should collect the terms with the same power of $e$ or, which is
equivalent, the same number of light fields. Then we should assign the
differential operators $\partial_t/M$ and $\nabla/M$ some power of the
parameter $v$ introduced in the previous section according to one of the power
counting schemes. The Lagrangian is then of the form
\begin{equation}
  \Lap = \bar\Lap^0 + \La_\lf^0 + \sum_{\mu,\nu}^\infty \left( \frac{e}{2M}
  \right)^\mu\Lap^{(\mu,\nu)}.
\end{equation}
Here, $\bar\Lap^0$ contains only the leading part of $\Lap^0$ in the parameter
$v$. The term $\Lap^{(\mu,\nu)}$ contains $\mu$ light fields and $\nu$ powers
of $v$.

In the HM scheme we formally assign
\begin{align}
  \frac{i\partial_t-M}{M} &= O(v)  \\
  \frac{\nabla}{M} &= O(v)
\end{align}
irrespective of what field they act on. The interaction independent pieces are
given by
\begin{align}
  \bar\Lap^0 &\equiv \Laphm \doteq \hfp^*\Dphm\hfp \\
  \Dphm &= i\partial_t - M \\
  \sum_{m=2}^\infty\Lap^{(0,m)} &= \hfp^*(M-\sqrt{M^2-\Delta})\hfp.
\end{align}
In the NR scheme we count time derivatives differently:
\begin{align}
  \frac{i\partial_t-M}{M} &= O(v^2) \;\;\text{when acting on a heavy field} \\
  \frac{i\partial_t}{M} &= O(v) \;\;\text{when acting on a light field} \\
  \frac{\nabla}{M} &= O(v) \;\;\text{always}
\end{align}
and
\begin{align}
  \bar\Lap^0 &\equiv \Lapnr \doteq  \hfp^*\Dpnr\hfp \\
  \Dpnr &= i\partial_t - M + \frac{\Delta}{2M} \\
  \sum_{m=2}^\infty\Lap^{(0,m)} &= \hfp^*(M - \frac{\Delta}{2M}
  -\sqrt{M^2-\Delta})\hfp. 
\end{align}
The leading terms of the interaction Lagrangians $\Lap^{(\mu,\nu)}$ for both
schemes are shown in table~\ref{tab:Lmunu}.

\renewcommand{\arraystretch}{1.6}    
\begin{table}[htbp]
  \begin{center}
    \begin{tabular}{|c|l|l|}
      \hline
      $(\mu,\nu)$ & HM & NR \\ \hline\hline 
      $(1,0)$ & $\hfp^*\lf\hfp$ 
      & $\hfp^*\lf\hfp$ \\ \hline 
      $(1,2)$ & $\frac{1}{4M^2}\hfp^*(\lf\Delta+\Delta\lf)\hfp$ 
      & $\frac{1}{4M^2}\hfp^*(\lf\Delta+\Delta\lf)\hfp$ \\ \hline 
      $(1,4)$ & $\frac{1}{16M^4}\hfp^*(\Delta\lf\Delta +
      \frac{5}{2}\lf\Delta^2$  
      & $\frac{1}{16M^4}\hfp^*(\Delta\lf\Delta +
      \frac{5}{2}\lf\Delta^2$  \\
      & $+\frac{5}{2}\Delta^2\lf)\hfp$ 
      & $+\frac{5}{2}\Delta^2\lf)\hfp$ \\ \hline
      $(2,0)$ & $\frac{1}{2M}\hfp^*\lf^2\hfp$ 
      & $\frac{1}{2M}\hfp^*\lf^2\hfp$ \\ \hline 
      $(2,1)$ & $-\frac{1}{4M^2}\hfp^*\lf(i\partial_t-M)\lf\hfp$ 
      & $\frac{-1}{4M^2}\hfp^*\lf(i\partial_t\lf)\hfp$ \\ \hline  
      $(2,2)$ & $\frac{1}{8M^3}\hfp^*(\lf^2\Delta+\Delta\lf^2]$ 
      & $\frac{1}{8M^3}\hfp^*(\lf^2\Delta+\Delta\lf^2 + 3\lf\Delta\lf$ \\
      & $+ 3\lf\Delta\lf+ \lf[i\partial_t-M]^2\lf)\hfp$ 
      & $- \lf(\partial_t^2\lf) -\frac{1}{2M}\lf^2(i\partial_t
      - M))\hfp$ \\ \hline
      $(3,0)$ & $\frac{1}{4M^2}\hfp^*\lf^3\hfp$ 
      & $\frac{1}{4M^2}\hfp^*\lf^3\hfp$  \\ \hline
    \end{tabular}
    \caption{The leading terms of the interaction Lagrangians
      $\Lap^{(\mu,\nu)}$. $\mu$ and $\nu$ denote the number of powers of $e$
      and $v$, respectively.}
    \label{tab:Lmunu}
  \end{center}
\end{table}

In the following, we first consider a free field and discuss the form of the
propagators to be used in perturbation theory. Then we formulate a power
counting for Green's functions to find out which vertices of the effective
Lagrangian must be used to calculate them to some order in $v$.  Finally, we
state how Green's functions can be calculated in a systematic way from the
generating functional.

\subsection{Free Propagators}
%----------------------------
The propagators to be used in perturbation theory are derived from the
Lagrangians $\Laphm$ and $\Lapnr$. In the notation of
appendix~\ref{app:notation} 
\begin{align}
  \eval{x}{\Dphm^{-1}}{y} &= \dphm(x-y) = \int\frac{d^4p}{(2\pi)^4}
  \frac{e^{-ip(x-y)}}{E+i\epsilon}  \\ 
  \eval{x}{\Dpnr^{-1}}{y} &= \dpnr(x-y) = - \int\frac{d^4p}{(2\pi)^4}
  \frac{e^{-ip(x-y)}}{\frac{\mathbf p^2}{2M}-E-i\epsilon},
\end{align}
where $E=p^0-M$. The operators in the Lagrangian $\Lap^{(0,\nu)}$ are
considered to be corrections to these lowest order propagators.  By resumming
insertions of $\mathbf p^2/2M$ in the HM propagator, we obtain the propagator
of the NR scheme.
\begin{align}
  \dphm(p)\left(1+\frac{\mathbf p^2}{2M}\frac{1}{E} + \left(\frac{\mathbf
  p^2}{2M}\frac{1}{E}\right)^2 + \dots\right) = \dpnr(p).
\end{align}
Similarly, by including higher and higher corrections and resumming them, we
recover the full propagator
\begin{equation}
  \Delta_+(p) = -\frac{1}{\omega(\mathbf p) - p^0 - i\epsilon}.
\end{equation}

\subsection{Naive Power Counting for Green's Functions}
%------------------------------------------------------
We would like to find a way how one can read off the power of $v$ to which a
certain Graph contributes. Every Graph can be characterized by the following
parameters

\begin{tabular}{ll}
  $E_\hf$ & $\#$ of external heavy lines \\
  $I_\hf$ & $\#$ of internal heavy lines \\
  $I_\lf$ & $\#$ of internal light lines \\
  $N_{\mu,\nu}$ & $\#$ of vertices with $\mu$ powers of $e$ and $\nu$ powers
  of $v$ \\ 
  $L$ & $\#$ of loops.
\end{tabular}

In addition, let $P$ denote the power of $1/v$ of the heavy propagator. We
have $P=1$ and $P=2$ in the HM and NR schemes, respectively.  Excluding
external lines, the total power $d$ of $v$ of the diagram is given by
\begin{equation}
  d = 4L - PI_\hf - 2I_\lf + \sum_{\mu,\nu} \nu N_{\mu,\nu}.
\end{equation}
Using the well known ``topological'' relations (the factor 2 in front of
$N_{\mu,\nu}$ is due to the fact that at each vertex exactly two heavy lines
meet) 
\begin{align}
  L &= I_\hf + I_\lf + 1 - \sum_{\mu,\nu} N_{\mu,\nu} \\
  E_\hf  &= \sum_{\mu,\nu}2N_{\mu,\nu} - 2I_\hf,
\end{align}
we get
\begin{equation}
  \label{eq:power-counting}
  d = 2(L+1) - \frac{2-P}{2} E_\hf + \sum_{\mu,\nu} N_{\mu,\nu}(\nu-P).
\end{equation}

This formula is certainly correct for $L=0$ because all factors of $v$ are
explicit and there are no integrations over internal momenta. Loops are
complicated functions of the external momenta and may produce additional
factors of $v$ which can upset this naive power counting. We will briefly come
back to this point below and consider only tree graphs for now.

Remember that we always work to a fixed order in the fundamental coupling in
$e$. Therefore, the sum $\sum_{\mu,\nu}\mu N_{\mu,\nu}$ must be the same for
every graph contributing to some Green's function. From
eq.~\eqref{eq:power-counting} we can see that the leading contribution is
given by the graph with as few powers of $v$ as possible. Corrections 
can be systematically obtained by including vertices with more powers of $v$.

\subsection{Perturbation Series}
%-------------------------------
We are now in a position to formulate how a Green's function can be
calculated from the generating functional
\begin{align}
  Z[j,j^*,J] &= \frac{1}{\mathcal Z}\int [dl][d\hfp][d\hfp^*] e^{iS_+ + \int
  j^*\hfp + \hfp^* j + J\lf} \\
\mathcal Z &= \int [dl][d\hfp][d\hfp^*] e^{iS_+} \\
S_+ &= \int d^4x \Lap(x).
\end{align}
The first step towards the perturbation theory is the separation of the action
$S_+$ into a ``free'' part (which must be quadratic in the field) and an
``interacting'' part
\begin{equation}
  S_+ = S_+^0 + S_+^{\text{int}}.
\end{equation}
This decomposition depends on the counting scheme and we set
\begin{align}
  S_{+,\text{HM,NR}}^0 &= \int d^4x \La_{+,\text{HM,NR}}^0 \\
  S_{+,\text{HM,NR}}^{\text{int}} &= S_+ - S_{+,\text{HM,NR}}^0.
\end{align}

The Gaussian average of some functional $F$ of the fields $\hfp$ and $\lf$ is
denoted by
\begin{equation}
  \gav{F[\hfp,\hfp^*,\lf]}^{\text{HM,NR}} \doteq \frac{\int
  [d\lf][d\hfp][d\hfp^*] 
  F[\hfp,\hfp^*,\lf] e^{iS_{+,\text{HM,NR}}^0}}{\int
  [d\lf][d\hfp][d\hfp^*]e^{iS_{+,\text{HM,NR}}^0}}. 
\end{equation}
In particular, the free propagators are given by
\begin{align}
  i\dphm(x-y) &= \gav{\hfp(x)\hfp^*(y)}^{\text{HM}} \\
  i\dpnr(x-y) &= \gav{\hfp(x)\hfp^*(y)}^{\text{NR}} \\
  \frac{1}{i}\Delta_m(x-y) &= \gav{\lf(x)\lf(y)}.
\end{align}
The latter is the same in both schemes.  In the notation set up in
section~\ref{sec:p-q-greens-functions}, the connected $n$-point functions are
written as (we should put indices HM or NR here as well but we suppress them in
order to simplify the notation)
\begin{equation}
  G_+^{(a,b)}(x,y,z) =
  \gav{\hat\hfp(x)\hat\hfp^*(y)\hat\lf(z)e^{iS_+^{\text{int}}}}_c. 
\end{equation}
The perturbation series is obtained by expanding the exponential in powers of
$v$ with the constraint that $\sum_{\mu,\nu}\mu N_{\mu,\nu}$ is fixed (see
above). After the expansion we are left with Gaussian integrals which can be
reduced to sums of products of propagators owing to the Wick theorem.

%%% Local Variables: 
%%% mode: latex
%%% TeX-master: "~/diss/tex/diss"
%%% End: 
\section{Compton Scattering at Tree Level}
%-----------------------------------------
Let us calculate the tree level truncated Green's function $G_{+,tr}^{(1,2)}$
in the HM scheme to next-to-next-to leading order. Applying
formula~\eqref{eq:power-counting}, we find the combinations of vertices that
yield a specific power of $v$ displayed in table~\ref{tab:compton-vertices}
\begin{table}[htbp]
  \begin{center}
    \begin{tabular}{|c|l|}
      \hline
      $d$ & Vertices \\ \hline\hline
      -1 & $N_{1,0}=2$ \\ \hline
      0  & $N_{2,0}=1;\; N_{1,0}=2,N_{0,2}=1$ \\ \hline
      1  & $N_{2,1}=1;\; N_{1,0}=2,N_{0,2}=2;\; N_{1,2}=1,N_{1,0}=1$ \\ \hline
    \end{tabular}
    \caption{The combination of vertices that yield a certain power $d$ of $v$
      for Compton scattering in the HM scheme.}
    \label{tab:compton-vertices}
  \end{center}
\end{table}
We can see, for example, that the leading term is of $O(1/v)$ (it is just the
propagator $\dphm$) and consists of two vertices of the Lagrangian
$\Lap^{(1,0)}$. At $O(v^0)$, we can either use one vertex from $\Lap^{(2,0)}$
or two from $\Lap^{(1,0)}$ together with $\Lap^{(0,2)}$, which is an insertion
of $\mathbf p^2/2M$.

The result is
\begin{align}
  \frac{1}{i}G_{+,tr}^{(1,2)} =&\; \frac{-1}{4M^2}\frac{1}{E_q + k_1^0} \left(
  1 + 
  \frac{(\mathbf q + \mathbf k_1)^2}{2M(E_q+k_1^0)} \right. \\
  &\;+ \left. \frac{(\mathbf q
  + \mathbf k_1)^4}{4M^2(E_q+k_1^0)^2} - \frac{\mathbf q^2 + \mathbf p^2 +
  2(\mathbf q + \mathbf k_1)^2}{4M^2} + O(v^3) \right) \\
  &\;+ \frac{1}{8M^3}\left( 1 - \frac{E_q+k_1^0}{2M} + O(v^2) \right).
\end{align}
According to the truncation rule given in eq.~\eqref{eq:p-a-truncation-rule},
we must multiply with
\begin{equation}
  \sqrt{2\omega(\mathbf q)}\sqrt{2\omega(\mathbf p)} = 2M\left( 1 +
  \frac{\mathbf q^2 + \mathbf p^2}{4M^2} + O(v^4) \right)
\end{equation}
to compare with the truncated Greens function $G_{tr}^{(1,2)}$ of the
relativistic theory. Comparing the result with
eq.~\eqref{eq:Gtr12-HM-expanded}, we see that 
\begin{equation}
  G_{tr}^{(1,2)} = \bar G_{+,tr}^{(1,2)}
\end{equation}
is true to $O(e^2 v^2)$. The amplitude $B_s^{(2)}$ for Compton scattering
obtained from $G_{+,tr}^{(1,2)}$ is therefore the same as the one in the
relativistic theory with the same precision.

%%% Local Variables: 
%%% mode: latex
%%% TeX-master: "~/diss/tex/diss"
%%% End: 
\section{Power Counting Beyond Tree-Level}
%-----------------------------------------
\label{sec:beyond-tree}
Consider the contribution of the graph
\begin{displaymath}
  \includegraphics[scale=0.8]{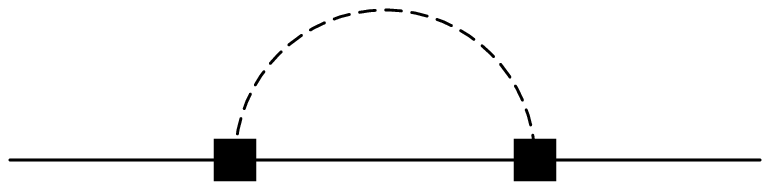}
\end{displaymath}
to the self energy of the heavy particle, where the boxes represent vertices
from $\Lap^{(1,2)}$, i.e. they have two powers of $v$. According to the
formula given in eq.~\eqref{eq:power-counting}, this diagram is of $O(v^5)$ in
the HM scheme and of $O(v^4)$ in NR. If $p$ is the momentum that flows through
the diagram, the loop is a function that depends only on $E=p^0-M$ and $m$ in
the HM scheme but on $E,m$ and $\mathbf p$ in the NR scheme. 
The integrals are of the form
\begin{align}
  I_{\text{HM}} &= I_{\text{HM}}\left(\frac{E}{m}\right) \\
  I_{\text{NR}} &= I_{\text{NR}}\left(\frac{E}{m},\frac{\mathbf
  p}{m},\frac{m}{M}\right). 
\end{align}
The argument of $I_{\text{HM}}$ is of $O(1)$ but the one of $I_{\text{NR}}$
contains a part that is of $O(v)$. Therefore, the loop destroys the naive
power counting in the NR scheme\footnote{This actually depends on the
  regularization prescription. If one uses a momentum space cutoff
  $M\alpha\ll\Lambda\ll M$, the loop starts contributing at the naive order
  see, for example, ref.~\cite{kinoshita-nio-1}. In dimensional
  regularization, however, this is not true}: unless the integral does
not really depend on $m/M$ by chance, it produces factors of $v$ which either
raise or lower the naive power of $v$. The former would not be so bad but the
latter is a disaster because one must expect that {\em all} loop graphs start
contributing at lowest order.

There is, however, a scenario, where this catastrophe is
reduced to a mere inconvenience. If the terms that contribute to a lower order
than the naive one are such that they can be absorbed in the coupling
constants of the Lagrangian (i.e. polynomials in the energies and momenta),
systematic perturbation theory is still possible, because only a
finite number of graphs contribute to the ``interesting'' (non-polynomial)
part of the Green's function. The inconvenience is that whenever one pushes
the calculation to the next higher order one has to re-match the effective
coupling constants (either to the fundamental theory, if possible, or directly
to experiment). 

It is believed that this is indeed what happens and was checked in an explicit
one-loop calculation~\cite{1-loop-renormalization}. 

In the HM scheme, the problem is absent\footnote{As a side remark, it is
  interesting to note that this fact is the reason why HBCHPT was introduced
  to replace the original relativistic treatment of the pion-nucleon
  system~\cite{gasser-sainio-svarc}, which suffers from the same power
  counting problem. It was recently shown that a new regularization scheme
  restores power counting in the manifestly relativistic
  formulation~\cite{becher-leutwyler}}. However, as mentioned in the
introduction, this scheme is not suited for systems where two heavy particles
can form a bound state because it leads to spurious infrared divergences, which
vanish only upon a resummation of certain contributions (see for
example~\cite{eichten-hill1}).

%%% Local Variables: 
%%% mode: latex
%%% TeX-master: "~/diss/tex/diss"
%%% End: 
\chapter{Summary and Outlook}
%============================
\markboth{SUMMARY AND OUTLOOK}{} 
In this work, we have investigated
effective theories describing heavy and light scalar particles in the
low--energy regime.  First, we discussed the concept of the physical mass and
of the matching condition for S--matrix elements in a general setting, and
then proposed a matching procedure for off--shell Green's functions, that
leads -- due to different notions of one--particle irreducibility in the
original and effective theory -- to a specific truncation prescription in the
effective theory.

We then investigated these matching conditions for a Yukawa interaction
between two heavy ($\hf$) and one light field ($\lf$).  First, we treated the
light field as an external source and constructed two non-local Lagrangians
that are equivalent to the full theory in the pure particle- and anti-particle
sectors. Adding dynamics for the light field, we showed that the amplitudes
and properly truncated Green's functions of the effective Lagrangians indeed
satisfy the proposed matching conditions to all orders in the coupling in the
presence of a UV regulator.

In order to arrive at a local Lagrangian, we discussed the $1/M$ expansion of
tree-level scattering.  We classified all physical processes by the number of
heavy particles and anti-particles in the initial and final states,
distinguishing
\begin{itemize}
\item {\it soft} processes: initial and final states contain only heavy
  particles or only heavy anti particles, like 
\begin{displaymath}
  \hf\lf\rightarrow\hf\lf\lf, 
\end{displaymath}
\item {\it semi-hard} processes: both types of particles are present, but
  their number is separately conserved, like 
\begin{displaymath}
  \hf \bar{\hf}\rightarrow \hf \bar{\hf}\lf\lf, 
\end{displaymath}
\item {\it hard} processes: number of particles and anti particles is not
  conserved separately, e.g., 
\begin{displaymath}
  \hf\bar{\hf}\rightarrow \lf\lf.  
\end{displaymath}
\end{itemize}
Starting from the nonlocal Lagrangian, we then constructed the effective local
Lagrangian for soft and semi--soft processes at low orders in the $1/M$
expansion.

Hard processes play a special role in this setting: their $1/M$ expansion is
difficult, because there is so much energy released that some light particles
may become very hard, while others stay soft. Neither did we find a
satisfactory treatment of these processes in the literature, nor can we offer
one at this moment\footnote{To mention an example, we consider the decay of Ortho-- or
  Parapositronium in the framework of nonrelativistic QED -- it requires
 the inclusion of the hard processes $e^+e^-\rightarrow
n\gamma$. In the literature, the problem is
  circumvented by use of a nonhermitean Lagrangian \cite{labelle-thesis}.
  While this may be useful as far as the calculational purpose is concerned,
  it is clear that there is room for improving this framework.}. Work
on the problem is in progress.

Extending the expansion to off-shell Green's functions, we found that there is
no natural way to count the energies of heavy particles relative to their
momenta (being no longer related through the on-shell condition). We
introduced a bookkeeping parameter $v$ and defined two possible counting
schemes by assigning powers of it to energies and momenta of the particles and
showed how -- in a systematic expansion in the fundamental coupling and in the
parameter $v$ -- tree-level Green's functions can be calculated.  We checked
the method in the case of Compton scattering.

The final aim of this programme is the application of effective
theories to the decay of bound states, like $\pi^+\pi^-\rightarrow
\pi^0\pi^0$, and to relate these processes to the underlying theory of strong
interactions. For this purpose, one needs to include hard processes in the
framework, and to set up a consistent and systematic power counting in the
scattering sector (including loops) as well as in the bound state calculation
where Rayleigh--Schr\"odinger perturbation theory may be applied. Finally, one
has to show how the effective Lagrangians describing QCD at low energies are
incorporated in order to arrive at the above described aim.  First steps in
this direction are already done
\cite{labelle-thesis,labelle-lepage,kinoshita-nio-1,
  labelle-buckley, kong-ravndall,labelle-retardation} or will soon be
completed \cite{energy-shift}.

%%% Local Variables: 
%%% mode: latex
%%% TeX-master: "diss"
%%% End: 
\thispagestyle{empty}
\cleardoublepage
\thispagestyle{myheadings}
\chapter*{Acknowledgements}

I would like to thank J\"urg Gasser for his advice, guidance and
encouragement and for passing his insight into physics on to me.

My thanks also go to Vito Antonelli for collaboration during his
stay in Bern and to Akaki Rusetsky for interesting discussions and his
hospitality during my brief stay in Dubna.

Last but not least I would like to thank all the people of the institute who
have contributed to this work in one way or another. I especially enjoyed the
many discussions with Thomas Becher and Markus Leibundgut, who were always
willing to stop working long enough to talk about life, physics and
the world in general.

%%% Local Variables: 
%%% mode: latex
%%% TeX-master: "diss"
%%% End: 
\begin{appendix}
  \markboth{APPENDIX}{}

\chapter{Notation}
%=================
\label{app:notation}
\subsubsection{Metric}
We work in Minkowski space with a signature of $(1,-1,-1,-1)$.  Three-vectors,
denoted by boldface letters, are the three-dimensional parts of 
contravariant four-vectors
\begin{equation}
  x^\mu = \{x^0,x^1,x^2,x^3\} = \{x^0, \mathbf x\}
\end{equation}
except for the three-dimensional gradient
\begin{equation}
  \mathbf\nabla = \{\partial_1, \partial_2, \partial_3\},
\end{equation}
where
\begin{equation}
  \partial_\mu \equiv \frac{\partial}{\partial x^\mu}.
\end{equation}

\subsubsection{Fourier Transform}
The Fourier transform $f(p)$ of a function $f(x)$ is defined by
\begin{equation}
  f(x) = \int\frac{d^4p}{(2\pi)^4} e^{-ipx}f(p).
\end{equation}

\subsubsection{Green's functions}
Let $\phi$ be a complex field and $x$ and $y$ denote sets
$(x_1,x_2,\dots,x_n)$, $(y_1,y_2,\dots,y_n)$ of coordinates. We use the
shorthand form
\begin{equation}
  \hat\phi(x) \doteq \phi(x_1)\phi(x_2)\dots\phi(x_n).
\end{equation}
The vacuum expectation value of the time ordered product of fields is written
as
\begin{equation}
  G (x,y) = \eval{0}{T\hat\phi(x)\hat\phi^\dag(y)}{0}.
\end{equation}
Assuming translation invariance, the Fourier transform of $G$ is defined
by 
\begin{equation}
  (2\pi)^4\delta^4(P-Q)G (p,q) = 
  \int d^4x d^4y e^{i\sum_{i=1}^n(p_ix_i-q_iy_i)}G(x,y),
\end{equation}
where $P=\sum_{i=1}^n p_i$ and $Q=\sum_{i=1}^n q_i$. With this convention, the
$p_i$ and $q_i$ denote the physical momenta of outgoing and incoming
particles if we let the time components $x_i^0$ and $y_i^0$ tend to $+\infty$
and $-\infty$, respectively. 

In the case of a real scalar field $\varphi$, we define 
\begin{equation}
  G(x) = \eval{0}{T\hat\varphi(x)}{0}
\end{equation}
and ($K=\sum_{i=1}^n k_i$)
\begin{equation}
  (2\pi)^4\delta^4(K)G(k) = 
  \int d^4x e^{i\sum_{i=1}^nk_ix_i}G(x).
\end{equation}
Here, the momenta $k_i$ correspond to outgoing particles in the same sense as above.

\subsubsection{Operators}
Let $\mathbf O$ be an operator that acts in some Hilbert space $\mathcal H$ of
functions defined in Minkowski space.  In Dirac notation, the orthogonality
and closure relations for the $x$ basis read
\begin{align}
  \sprod{x}{y} &= \delta^4(x-y) \\
  \int d^4x \ket{x}\bra{x} &= \mathbf 1.
\end{align}
The $x$ representations of $f\in\mathcal H$ and $\mathbf O$ are denoted by
\begin{align}
   f(x) &= \sprod{x}{f} \\
   O(x,y) &= \eval{x}{\mathbf O}{y}.
\end{align}
Accordingly, the action of $\mathbf O$ on $f$ reads
\begin{equation}
  (\mathbf O f)(x) = \int d^4 y O(x,y)f(y).
\end{equation}
 A differential
operator $\mathbf D$ has the representation
\begin{equation}
  \eval{x}{\mathbf D}{y} = \delta^4(x-y)D_y
\end{equation}
so that
\begin{equation}
  (\mathbf D f)(x) = D_x f(x).
\end{equation}
For any translation invariant operator, i.e.  $\eval{x}{\mathbf O}{y} =
O(x-y)$, we have
\begin{equation}
  \Box_x O(x-y) = \Box_yO(x-y).
\end{equation}
If $\mathbf D$ is an invariant differential operator (i.e. a function of
$\Box$) and $\mathbf O$ translation invariant, one may check, using partial
integration, that
\begin{equation}
  \label{eq:rot-inv-diff-op}
  (\mathbf D\mathbf O f)(x) = (\mathbf O\mathbf D f)(x).
\end{equation}

%%% Local Variables: 
%%% mode: latex
%%% TeX-master: "~/diss/tex/diss"
%%% End: 
\chapter{Klein-Gordon Green's Functions}
%---------------------------------------
\label{app:KG}
A Green's function $G(x)$ of the Klein-Gordon equation is defined by
\begin{equation}
  \label{eq:KG-gf-def}
  D_MG(x)\doteq (\Box + M^2)G(x) = \delta^4(x)
\end{equation}
together with some boundary conditions. The solution that is a superposition
of incoming plane waves for $x_0<0$ and of outgoing plane waves for $x_0>0$ is
the Feynman propagator
\begin{align}
  \label{eq:feynman-prop}
  \Delta_M(x) &= \int\frac{d^4p}{(2\pi)^4} \frac{e^{-ipx}}{M^2-p^2-i\epsilon}
  \notag \\ 
  &= i\int
  \frac{d^3p}{(2\pi)^3 2\omega(\mathbf p)}e^{i\mathbf p\cdot\mathbf x} \left(
  \theta(x^0)e^{-i\omega(\mathbf p)x^0} + \theta(-x^0) e^{i\omega(\mathbf
  p)x^0}\right),
\end{align}
where $\omega(\mathbf p) = \sqrt{M^2+\mathbf p^2}$.

The Klein-Gordon operator $D_M$ can be decomposed into two first order
differential operators 
\begin{eqnarray}
  \label{eq:def-D+-}
  D_M &=& D_+ D_- \nonumber \\
  D_\pm &=& \pm i\partial_t - \sqrt{M^2-\Delta}.
\end{eqnarray}
The operator
\begin{equation}
  \label{eq:def-d}
  d = (2\sqrt{M^2-\Delta})^{-\frac{1}{2}}
\end{equation}
plays an important role in the construction of the non-relativistic
Lagrangian. Its action on a function $f$ is defined through the Fourier
representation
\begin{equation}
  (df)(x) = \int \frac{d^4p}{(2\pi)^4} \frac{f(p)}{\sqrt{2\omega(\mathbf p)}}
  e^{-ipx}. 
\end{equation}

The functions
\begin{align}
  \label{eq:d-pm}
  \Delta_\pm(x) &= - \int \frac{d^4p}{(2\pi)^4} \frac{e^{-ipx}}{\omega(\mathbf
  p)
  \mp p^0 - i \epsilon}  \notag \\
&= -i\theta(\pm x^0) \int\frac{d^3p}{(2\pi)^3} e^{\mp i\omega(\mathbf p)x^0 +
  i\mathbf p\cdot\mathbf x}
\end{align}
are Green's functions of $D_\pm$, i.e. 
\begin{equation}
  D_\pm\Delta_\pm(x) = \delta^4(x)
\end{equation}
and the boundary conditions are chosen such that $\Delta_+(x)=0$ for $x^0<0$
and $\Delta_-(x)=0$ for $x^0>0$. Comparing (\ref{eq:d-pm}) with
(\ref{eq:feynman-prop}) we find
\begin{equation}
  \label{eq:fp-as-d_pm}
  \Delta_M(x) = -d^2(\Delta_+(x)+ \Delta_-(x)).
\end{equation}

The Green's functions can be viewed as the inverse of the corresponding
differential operators. In the notation introduced in
appendix~\ref{app:notation}, we write
\begin{align}
  \eval{x}{D_M^{-1}}{y} &= \Delta_M(x-y) \\
  \eval{x}{D_\pm^{-1}}{y} &= \Delta_\pm(x-y).
\end{align}
In operator notation, eq.~\eqref{eq:fp-as-d_pm} can be written in any of the
forms (cf. eq.~\eqref{eq:rot-inv-diff-op}) 
\begin{eqnarray}
  \label{eq:DM-as-Dpm}
  D_M^{-1} &=& -d^2(D_+^{-1} + D_-^{-1}) = -(D_+^{-1} +
  D_-^{-1})d^2 \notag \\
  &=& -d(D_+^{-1} + D_-^{-1})d.
\end{eqnarray}

Finally, with the convention of appendix~\ref{app:notation}, the Fourier
transforms are given by 
\begin{align}
  \Delta_M(p) &= \frac{1}{M^2-p^2-i\epsilon} \\
  \Delta_\pm(p) &= -\frac{1}{\omega(\mathbf p)\mp p^0-i\epsilon}.
\end{align}

%%% Local Variables: 
%%% mode: latex
%%% TeX-master: "~/diss/tex/diss"
%%% End: 
\chapter{Canonical Quantization of Free Fields}
%==============================================
\label{app:canonical-quantisation}
Let us briefly recall the canonical quantization procedure for a complex
scalar field with the Lagrangian
\begin{equation}
  \La^0 = \partial_\mu\hf^*\partial^\mu\hf - M^2\hf^*\hf.
\end{equation}
The conjugate field is defined by
\begin{equation}
  \pi(t,\mathbf x) = \frac{\partial \La^0}{\partial\dot\hf(t,\mathbf x)} =
  \dot\hf^*(t,\mathbf x) 
\end{equation}
where $\dot\hf(t,\mathbf x) = \partial_0\hf(t,\mathbf x)$. 
The only non-vanishing Poisson bracket is
\begin{equation}
  \label{eq:rel-poisson}
  \{\hf(t,\mathbf x), \pi(t, \mathbf y)\} = \delta^3(\mathbf x-\mathbf y)
\end{equation}
and the most general solution of the equation of motion
\begin{equation}
  (\Box + M^2)\hf(x) = 0
\end{equation}
is a superposition of plane waves
\begin{equation}
  \hf(x) = \int d\mu(p)\left( a(\mathbf p)e^{-ipx} + b^*(\mathbf p)e^{ipx}
  \right),
\end{equation}
where the invariant measure is defined by
\begin{equation}
  d\mu(p) = \frac{d^3p}{(2\pi)^32p^0}
\end{equation}
and the momentum is on the mass shell
\begin{equation}
  p^0 = \omega(\mathbf p) = \sqrt{M^2+\mathbf p^2}.
\end{equation}
The factor $(2\pi)^3$ is conventional and is chosen for later convenience.
Quantization is performed by replacing $\hf$ and $\pi$ by operators which
satisfy the equal-time commutation relation
\begin{equation}
  [\hf(t,\mathbf x), \pi(t, \mathbf y)] = i\delta^3(\mathbf x-\mathbf y). 
\end{equation}
The coefficient functions $a$ and $b$ are also operators and obey
\begin{equation}
  [a(\mathbf p), a^\dag(\mathbf q)] = [b(\mathbf p), b^\dag(\mathbf q)] = 
  2\omega(\mathbf p)(2\pi)^3\delta^3(\mathbf p-\mathbf q).
\end{equation}
The operators $a^\dag$ and $b^\dag$ can be shown to create one-particle states
out of the vacuum
\begin{align}
  \ket{p} &= a^\dag(\mathbf p)\ket{0} \\
  \ket{\bar p} &= b^\dag(\mathbf p)\ket{0}.
\end{align}
We shall refer to them as particle and anti-particle states, respectively. The
vacuum contains by definition no particles and is defined by the conditions
\begin{align}
  a(\mathbf p)\ket{0} &= b(\mathbf p)\ket{0} = 0 \notag \\
  \sprod{0}{0} &= 1.
\end{align}
With these conventions, the states are normalized by
\begin{equation}
  \sprod{p}{q} = \sprod{\bar p}{\bar q} = 2\omega(\mathbf
  p)(2\pi)^3\delta^3(\mathbf p-\mathbf q). 
\end{equation}

Let us apply this formalism to the Lagrangians
\begin{equation}
  \La_\pm^0 = \hf_\pm^*(\pm i\partial_t - \sqrt{M^2-\Delta})\hf_\pm.
\end{equation}
The conjugate fields are
\begin{align}
  \pi_\pm(t,\mathbf x) = \frac{\partial\La_\pm}{\partial \dot\hf_\pm} = \pm
  i\hf_\pm^*
\end{align}
and the Poisson brackets are analogous to eq.~\ref{eq:rel-poisson}. The most
general solutions of the equations of motion
\begin{equation}
  (\pm i\partial_t-\sqrt{M^2-\Delta})\hf_\pm(x) = 0
\end{equation}
are
\begin{align}
  \hfp(x) &= \int d\bar\mu(p) a(\mathbf p) e^{-ipx} \\
  \hfm(x) &= \int d\bar\mu(p) b^*(\mathbf p) e^{ipx}
\end{align}
with $p^0=\omega(\mathbf p)$. This time, we chose the measure to be
\begin{equation}
  d\bar\mu(p) = \frac{d^3p}{(2\pi)^3}.
\end{equation}
Replacing the functions by operators, we find 
\begin{equation}
  [a(\mathbf p), a^\dag(\mathbf q)] = [b(\mathbf p), b^\dag(\mathbf q)] = 
  (2\pi)^3\delta^3(\mathbf p-\mathbf q).
\end{equation}
They create and destroy particles in the same way as described above. The only
difference is the normalization of these states, which now reads
\begin{equation}
  \sprod{p}{q} = \sprod{\bar p}{\bar q} = (2\pi)^3\delta^3(\mathbf p-\mathbf
  q).  
\end{equation}
We have sacrificed the rule to label different objects by different symbols to
simplify the notation. 

Note that the Fock spaces of the theories defined by $\La^0$ and $\Lap+\Lam$
are identical.

\chapter{Two-Point Functions}
%----------------------------
\label{app:2pf}
In section~\ref{sec:transition-amplitudes}, we consider transition amplitudes
in the two models
\begin{align}
  \tag{\ref{eq:L-gener}} \La &= \partial_\mu\hf^*\partial^\mu\hf-M^2\hf^*\hf
  + \bar\La^0 + \La^{\text{int}} \\
  \tag{\ref{eq:Lp-gener}} \Lap &= \hfp^*(i\partial_t-\sqrt{M^2-\Delta})\hfp +
  \bar\La^0 + \Lap^{\text{int}}
\end{align}
and show how they can be matched. For this procedure to work, it is necessary
that there exists an unambiguous definition of the physical mass of the heavy
particle. We must therefore examine the properties of the two-point functions
\begin{align}
  G(p) &= \int d^4x e^{ipx}\eval{0}{T \hf(x)\hf^\dag(0)}{0} \\
  \label{eq:Gp-def}
  G_+(p) &= \int d^4x e^{ipx}\eval{0}{T \hfp(x)\hfp^\dag(0)}{0}.
\end{align}
It is convenient to express them in terms of one-particle irreducible
functions\footnote{They can be obtained in perturbation theory from the
  Legendre transform of the generating functional of connected Green's
  functions} $\Sigma$,~$\Sigma_+$
\begin{align}
  G(p) &= \frac{1}{i}\frac{1}{M^2-p^2 + i\Sigma(p^2) -i\epsilon} \\
  \label{eq:Gp}
  G_+(p) &= \frac{1}{i}\frac{1}{\omega(\mathbf p)-p^0 + i\Sigma_+(p^0,\mathbf
    p^2) - i\epsilon},
\end{align}
with $\omega(\mathbf p) = \sqrt{M^2+\mathbf p^2}$. In the absence of
interactions, they reduce to the free propagators
\begin{align}
  G(p) &= \frac{1}{i}\Delta_M(p) = \frac{1}{i}\frac{1}{M^2-p^2-i\epsilon} \\
  G_+(p) &= i\Delta_+(p) = \frac{1}{i} \frac{1}{\omega(\mathbf p)- p^0
    -i\epsilon}
\end{align}
discussed in appendix~\ref{app:KG}.

The physical mass $\Mp$ is defined as the location of the pole of $G$
\begin{equation}
  \label{eq:Mp-def}
  M_{\text p}^2 = M^2 + i\Sigma(M_{\text p}^2)
\end{equation}
and we can write
\begin{equation}
  \label{eq:G2-phys}
  G(p) = \frac{1}{i}\frac{Z_\hf}{M_{\text p}^2-p^2-i\epsilon} +
  \text{regular},p^2\rightarrow\Mp^2,
\end{equation}
where the residue is given by
\begin{equation}
  \label{eq:Z-def}
  Z_\hf^{-1} = 1 - i\Sigma'(M_{\text p}^2).
\end{equation}
Let us focus on the pole at $p^0=\op(\mathbf p) = \sqrt{\Mp^2+\mathbf p^2}$
\begin{equation}
  \label{eq:G-pole-p0}
  G(p) = \frac{1}{i}\frac{1}{2\op(\mathbf p)}\frac{Z_\hf}{\op(\mathbf p)-p^0 -
  i\epsilon} + \text{regular},p^0\rightarrow\op(\mathbf p).
\end{equation}
We may isolate it in a different way by first writing $G$ as
\begin{equation}
  \label{eq:G-in-S}
  G(p) = \frac{1}{i}\Delta_M(p)\left(1 + S(p)\frac{1}{i}\Delta_M(p)\right),
\end{equation}
with
\begin{equation}
  S(p) = \frac{\Sigma(p^2)}{1+\Delta_M(p)i\Sigma(p^2)}.
\end{equation}
The r.h.s. sums up products of propagators with insertions of $\Sigma$. This
representation shows that the latter is really the 1-particle irreducible
two-point function with respect to $\Delta_M$. The idea is to define a new
irreducible function {\em with respect to} $\Delta_+$. It is clear that
$\Sigma$ is still irreducible in this new sense. In appendix~\ref{app:KG} it
is shown that
\begin{equation}
  \Delta_M(p) = -\frac{1}{2\omega(\mathbf p)}\left( \Delta_+(p) + \Delta_-(p)
  \right), 
\end{equation}
where $\Delta_-(p)=-1/(\omega(\mathbf p) + p^0)$, and we find that $S$
contains new irreducible functions, namely those obtained by connecting
factors of $\Sigma$ with $\Delta_-$, which is considered to be irreducible.
Therefore,
\begin{align}
  \label{eq:Sp}
  \hat\Sigma_+(p^0,\mathbf p^2) &\doteq \frac{\Sigma(p^2)}{2\omega(\mathbf p)}
  + 
  \frac{\Sigma(p^2)}{2\omega(\mathbf p)}i\Delta_-(p)
  \frac{\Sigma(p^2)}{2\omega(\mathbf p)} + \dots \notag \\
  &= \frac{\Sigma(p^2)}{2\omega(\mathbf p)-\Delta_-(p) i\Sigma(p^2)},
\end{align}
is the fundamental irreducible function with respect to $\Delta_+$. One may
easily check that in terms of $\hat\Sigma_+$, $S$ can be written as
\begin{equation}
  S(p) = \frac{2\omega(\mathbf p)\hat \Sigma_+(p^0,\mathbf p^2)}{1-\Delta_+(p)
  i\hat \Sigma_+(p^0,\mathbf p^2)}.
\end{equation}
Let us also define
\begin{align}
  \hat G_+(p) &\doteq i\Delta_+(p)\left( 1 + \frac{S(p)}{2\omega(\mathbf p)}
    i\Delta_+(p) \right) \\
  &= \frac{1}{i}\frac{1}{\omega(\mathbf p)-p^0 + i\hat \Sigma_+(p^0,\mathbf
    p^2) - i\epsilon}.
\end{align}
The equation
\begin{equation}
  \label{eq:Mpp-def}
  \op(\mathbf p) = \omega(\mathbf p) + i\hat\Sigma_+(\op(\mathbf p^2), \mathbf
  p), 
\end{equation}
which defines the location of the pole of $\hat G_+$, is equivalent
to~\eqref{eq:Mp-def} and we can write
\begin{equation}
  \label{eq:Gp-pole}
  \hat G_+(p) = \frac{1}{i}\frac{\hat Z_+(\mathbf p^2)}{\op(\mathbf p) - p^0 -
  i\epsilon} + \text{regular},p^0\rightarrow\op(\mathbf p),
\end{equation}
where
\begin{equation}
  \hat Z_+^{-1}(\mathbf p^2) = 1 - i\hat\Sigma_+'(\op(\mathbf p),\mathbf p^2)
\end{equation}
and the prime refers to the derivative with respect to $p^0$. With a little
algebra, we can cast eq.~\eqref{eq:G-in-S} into the form
\begin{align}
  G(p) =&\; \left(1-\frac{i\hat\Sigma_+(p^0, \mathbf p^2)}{\omega(\mathbf
      p)+p^0} \right)
  \frac{\hat G_+(p)}{\omega(\mathbf p)+p^0} \notag \\
  =&\; \frac{1}{i} \frac{1}{\omega(\mathbf p)+p^0} \left(1-\frac{i\hat
      \Sigma_+(p^0, \mathbf p^2)}{\omega(\mathbf p)+p^0} \right)
  \frac{\hat Z_+(\mathbf
    p^2)}{\op(\mathbf p) - p^0 - i\epsilon} \notag \\
  &\; + \text{regular},p^0\rightarrow\op(\mathbf p),
\end{align}
which is to be compared with eq.~\eqref{eq:G-pole-p0}. The relation between
the residues can be read off to be
\begin{equation}
  \label{eq:Z-Zp}
  \hat Z_+(\mathbf p^2) = \frac{(\omega(\mathbf p)+ \op(\mathbf p))^2} {4\op(\mathbf p)\omega(\mathbf p)}Z_\hf.
\end{equation}
Note that the $\mathbf p$ dependence of $\hat Z_+$ is entirely due to loop
corrections. At tree-level, where $\op(\mathbf p)=\omega(\mathbf p)$, the
residues are, of course, both equal to $1$.

Let us now come to the original $G_+$ defined in eq.~\eqref{eq:Gp-def}. From
the previous analysis we find that if we match the irreducible function
$\Sigma_+$ defined in eq.~\eqref{eq:Gp} to the full theory according to
\begin{equation}
  \Sigma_+(p^0,\mathbf p^2) = \hat\Sigma_+(p^0,\mathbf p^2),
\end{equation}
we also have $G_+(p)=\hat G_+(p)$ and the physical mass defined through
eq.~\eqref{eq:Mpp-def} is the same as the one in the relativistic theory. The
residues are related by eq.~\eqref{eq:Z-Zp}. Note that the matching is done
off-shell. The only relevant objects for physical quantities are the location
of the pole, defining the physical mass in terms of the parameters of the
theory, and its residue, providing the effective normalization of the field.
Any off-shell matching that does not change these properties is allowed. The
construction presented here singles out one of these possibilities rather
naturally.

To use these results in a calculation of physical quantities, we must
renormalize the theories so that they yield finite results when the regulator
is removed. The necessary counter terms at one-loop order are determined in
appendix~\ref{app:1-loop-renormalization}. The statements derived here can be
verified explicitly to this order in perturbation theory.

%%% Local Variables: 
%%% mode: latex
%%% TeX-master: "~/diss/tex/diss"
%%% End: 
\chapter{Reduction Formulae}
%===========================
\label{app:reduction}
The reduction formula gives the relationship between the residues of certain
Green's functions and physical scattering amplitudes.  The underlying
assumption is that particles involved in a scattering process behave like free
particles long before and long after the collision. This is called the
asymptotic condition and must be formulated carefully, see for example
\cite{zuber,bogoliubov}.

We first give a review of the facts in a relativistic theory and then consider
an effective theory which is not manifestly Lorentz invariant.

\section{Relativistic Theory}
%----------------------------
We consider the generic Lagrangian
\begin{align}
  \La &= \La^0+ \bar\La^0 + \La^{\text{int}} \notag \\
  \La^0 &= \partial_\mu\hf^*\partial^\mu\hf-M^2\hf^*\hf
\end{align}
introduced in section~\ref{sec:transition-amplitudes}. In the notation of
appendix~\ref{app:notation}, connected Green's functions of the heavy field
are denoted by
\begin{equation}
  \label{eq:red-def-G}
  G^{(2n)}(x,y) = \eval{0}{T\hat\hf(x)\hat\hf^\dag(y)}{0}_c.
\end{equation}
Canonical quantization of the free $\hf$ field leads to creation and
annihilation operators of one particle states as described in
appendix~\ref{app:canonical-quantisation}. The asymptotic condition says that
the interacting field behaves like a free field in the remote past and future
in the weak sense (only for matrix elements)
\begin{align}
  \hf(x) &\stackrel{x^0\rightarrow -\infty}{\rightarrow} Z_\hf^\frac{1}{2}
  \hfin(x) \\
  \hf(x) &\stackrel{x^0\rightarrow +\infty}{\rightarrow} Z_\hf^\frac{1}{2}
  \hfout(x)
\end{align}
The fields $\hfin$, $\hfout$ have all the properties of free fields and
Lorentz invariance implies that $Z_\hf$ is a constant, which is given by the
residue of the full two-point function
\begin{align}
  \label{eq:G-rel}
  G^{(2)}(p) &= \int d^4x e^{ipx}\eval{0}{T\hf(x)\hf^\dag(0)}{0} =
  \frac{1}{i}\frac{1}{M^2-p^2+i\Sigma(p^2)} \notag \\
  &= \frac{1}{i}\frac{Z_\hf}{M_{\text p}^2-p^2-i\epsilon} + \dots
\end{align}
The physical mass $\Mp$ and the residue $Z_\hf$ are defined through the
one-particle irreducible function $\Sigma$ by (see also
appendix~\ref{app:2pf})
\begin{align}
  \label{eq:Mp-rel}
  \Mp^2 &= M^2+i\Sigma(\Mp^2) \\
  \label{eq:Z-rel}
  Z_\hf^{-1} &= 1 - i\Sigma'(\Mp^2).
\end{align}
We define in- and out states by
\begin{align}
  \label{eq:in-state}
  \ketin{p} &= a_{\text{in}}^\dag(\mathbf p)\ket 0 \notag \\
  \ketout{p} &= a_{\text{out}}^\dag(\mathbf p)\ket 0
\end{align}
and similar for $\ketin{\bar p}$, $\ketout{\bar p}$. In fact, the Hilbert
spaces of in- and out states are identical and the scattering operator $S$ is
the isomorphism that maps a state $\ketin{i}$ into the space of out-states
\begin{equation}
  \ketin{i} = S\ketout{i}.
\end{equation}
Defining the $T$ operator by
\begin{equation}
  S = 1+iT,
\end{equation}
the amplitude to find the final state $\ketout{f}$ is given by
\begin{align}
  \label{eq:Tfi-rel}
  \sprod{f;\text{out}}{i;\text{in}} &=
  \sprod{f;\text{in}}{i;\text{in}}+i\eval{f}{T}{i} \notag \\ 
  &= \sprod{f;\text{in}}{i;\text{in}}+i(2\pi)^4\delta^4\left(P_f-P_i\right)
  T_{fi}, 
\end{align}
where we have also defined the $T$-matrix element $T_{fi}$.  If none of the
initial one-particle states are contained in the final state, the first term
on the r.h.s. vanishes.

Let's consider a configuration where there are $n$ heavy particles in the
initial and final states, i.e.
\begin{align}
  \ketin{i} &= \ketin{q_1,\dots,q_n} \\
  \ketout{f} &= \ketout{p_1,\dots,p_n}.
\end{align}
Reducing the in- and out states as described, for example, in \cite{zuber} we
find
\begin{multline}
  \eval{p_1,\dots,p_n;\text{in}}{S-1}{q_1,\dots,q_n;\text{in}}_c =
  \left(iZ_\hf^{-\frac{1}{2}}\right)^{2n}\int
  d^4xd^4ye^{i\sum_i(p_ix_i-q_iy_i)} \\
  (\Box_{x_1}+\Mp^2)\dots(\Box_{y_n}+\Mp^2)G^{(2n)}(x,y),
\end{multline}
where the subscript $c$ indicates that disconnected
contributions\footnote{There are no subsets of particles that do not interact}
are not included.  In terms of the truncated Green's function, this reads
\begin{multline}
  \eval{p_1,\dots,p_n;\text{in}}{S-1}{q_1,\dots,q_n;\text{in}}_c = \\
  (2\pi)^4\delta^4\left(P-Q\right) Z_\hf^n
  \left.G_{tr}^{(2n)}(p,q)\right|_{\text{on-shell}},
\end{multline}
where $P = \sum_i p_i, Q = \sum_i q_i$ and ``on-shell'' means
$p_i^0=\omega_{\text{p}}(\mathbf p_i) = \sqrt{\Mp^2+\mathbf p_i^2}$,
$q_i^0=\omega_{\text{p}}(\mathbf q_i)$. Finally, we read off the expression
for the $T$-matrix element for this process\footnote{The $T_{n\rightarrow n}$
  is not precisely the one defined in eq.~\eqref{eq:Tfi-rel} but only the
  contribution of the connected part}
\begin{equation}
  \label{eq:Tn->n}
  T_{n\rightarrow n} = \frac{1}{i}Z_\hf^n
  \left.G_{tr}^{(2n)}(p,q)\right|_{\text{on-shell}}. 
\end{equation}

\section{Effective Theory}
%----------------------------
Now consider the Lagrangian 
\begin{align}
  \Lap &= \Lap^0 + \bar\La^0 + \Lap^{\text{int}} \notag \\
  \La_\pm^0 = \hf_\pm^*D_\pm\hf_\pm
\end{align}
introduced in sections~\ref{sec:free-fields}
and~\ref{sec:eff-particle-theory}. The Green's functions $G_+^{(2n)}$ are
defined in analogy with eq.~\eqref{eq:red-def-G}.  Again, we start with the
quantization of the free $\hfp$ field as described in
appendix~\ref{app:canonical-quantisation}. It is very important that, up to
the normalization, the one-particle state $\ket{p}$ is the same as the one of
the relativistic theory: it describes a free scalar particle with momentum
$\mathbf p$ and energy $\sqrt{\Mp^2+\mathbf p^2}$.  Therefore, the Fock spaces
obtained by applying particle creation operators to the vacuum are
the same in both theories. Since the in and out states live in this Fock
space, the stage is set for an effective theory that can generate the same
transition amplitudes as a relativistic theory (cf.
sections~\ref{sec:eff-particle-theory} and~\ref{sec:amplitude-matching}).

Lorentz symmetry is not respected and the asymptotic condition for the
interacting field reads
\begin{align}
  \label{eq:H-asymp-+}
  \hfp(x) &\stackrel{x^0\rightarrow -\infty}{\rightarrow}
  Z_+(\Delta)^\frac{1}{2}\hfpin(x) \\ 
  \label{eq:H-asymp--}
  \hfp(x) &\stackrel{x^0\rightarrow +\infty}{\rightarrow}
  Z_+(\Delta)^\frac{1}{2}\hfpout(x).  
\end{align}
The symbol $Z_+(\Delta)$ represents a rotation invariant differential
operator. In momentum space, it becomes a function of $\mathbf p^2$ and, as in
the previous section, we expect it to be the residue of the full two-point
function, which, in terms of the irreducible part $\Sigma_+$, reads
\begin{equation}
  G_+^{(2)}(p) = \frac{1}{i}\frac{1}{\omega(\mathbf
  p)-p^0+i\Sigma_+(p^0,\mathbf p^2)-i\epsilon}. 
\end{equation}
It seems obvious to define the physical mass by the zero of the denominator
\begin{equation}
  \label{eq:Mp-eff}
  \op(\mathbf p) = \sqrt{\Mp^2+\mathbf p^2} = \omega(\mathbf
  p)+i\Sigma_+(\op(\mathbf p),\mathbf p^2). 
\end{equation}
However, for the most general rotation invariant Lagrangian, this would yield
a momentum-dependent object $\Mp$, which cannot serve as a mass parameter. In
view of our goal, which is to reproduce the scattering amplitudes of a fully
relativistic theory, we may impose a constraint on the interaction Lagrangian,
leading to a momentum-independent parameter $\Mp$, which can play the role of
the physical mass of the particle. This constraint can be easily implemented
in perturbation theory, where it results in a relation among the coupling
constants of the theory (see also the discussion in appendix~\ref{app:2pf}).
Assuming this is done, we can write
\begin{equation}
  \label{eq:G2-eff-phys}
  G_+^{(2)}(p) = \frac{1}{i}\frac{Z_+(\mathbf p^2)}{\op(\mathbf
  p)-p^0-i\epsilon} + \dots 
\end{equation}
The residue $Z_+$ is given by (the prime refers to the derivative with respect
to $p^0$),
\begin{equation}
  \label{eq:Z-eff}
  Z_+^{-1} = 1-i\Sigma_+'(\op(\mathbf p),\mathbf p^2).
\end{equation}
In- and out states are defined in analogy to eq.~\eqref{eq:in-state} and all
that was said about the $S$ and $T$ matrix in the previous section applies
also here.  The fact that $\hfp$ can only destroy a particle in the in-state
together with the hermiticity of the Lagrangian implies that the number of
heavy particles in the initial and final states must be the same. The
procedure of the reduction of in- and out-states can be applied to the
effective theory without any problems. The result is
\begin{multline}
  \eval{p_1,\dots,p_n}{S-1}{q_1,\dots,q_n}_c = \\
    i^{2n}\prod_{i=1}^nZ_+(\mathbf p_i^2)^{-\frac{1}{2}} Z_+(\mathbf
    q_i^2)^{-\frac{1}{2}} 
    \int d^4xd^4ye^{i\sum_i(p_ix_i-q_iy_i)} \\
    (\sqrt{\Mp^2-\Delta_{x_1}}-i\partial_{x_1^0})\dots
    (\sqrt{\Mp^2-\Delta_{y_n}}+i\partial_{y_n^0})G_+^{(2n)}(x,y)
\end{multline}
or, in terms of the truncated function,
\begin{multline}
    \eval{p_1,\dots,p_n}{S-1}{q_1,\dots,q_n}_c = 
    (2\pi)^4\delta^4\left(P-Q\right) \\  
    \prod_{i=1}^nZ_+(\mathbf p_i^2)^\frac{1}{2} Z_+(\mathbf
    q_i^2)^\frac{1}{2} \left.G_{+,tr}^{(2n)}(p,q)\right|_{\text{on-shell}}.
\end{multline}
The notion of on-shell is the same as before and the $T$ matrix element is
given by
\begin{equation}
  \label{eq:Tfi-eff}
  T_{n\rightarrow n}^+ = \frac{1}{i}\prod_{i=1}^nZ_+(\mathbf p_i^2)^\frac{1}{2}
    Z_+(\mathbf 
    q_i^2)^\frac{1}{2} 
  \left.G_{+,tr}^{(2n)}(p,q)\right|_{\text{on-shell}}. 
\end{equation}

%%% Local Variables: 
%%% mode: latex
%%% TeX-master: "~/diss/tex/diss"
%%% End: 
\chapter{Proof of eq.~\protect\eqref{eq:Tpm-T-equivalence}}
%==========================================================
\label{app:T}
\newcommand{\co}{\; \; ,} 
\newcommand{\scs}{\co \;}
\newcommand{\per}{ \; .}
We prove the decomposition for $G_1$ in
eq.~\eqref{eq:def-G1} -- the proof for the decompositions $G_{2,3,4}$ in that
equation is very similar.  Using
\begin{equation}
  A=D_++eB\scs C=D_- + eB\co\nonumber
\end{equation}
one has
\begin{equation}
  A-e^2BC^{-1}B=D_+ + eB-e^2BC^{-1}B\co\nonumber
\end{equation}
and
\begin{align}
  eB-e^2BC^{-1}B &= eBC^{-1}[C-eB]=eBC^{-1}D_- \notag \\
  &=eB(1+eD_-^{-1}B)^{-1}\co\nonumber
\end{align}
as a result of which $G_1$ becomes
\begin{equation}
  G_1
  =\left[1+eD_+^{-1}B
    (1+eD_-^{-1}B)^{-1}\right]^{-1}D_+^{-1}\per\nonumber
\end{equation}
With
\begin{equation}
  1+eD_+^{-1}B(1+eD_-^{-1}B)^{-1}=  \left[1+e(D_+^{-1}+D_-^{-1})B\right]
  (1+eD_-^{-1}B)^{-1}\co\nonumber
\end{equation}
we find
\begin{equation}
  G_1=(1+eD_-^{-1}B)\left[1+e(D_+^{-1}+D_-^{-1})B\right]^{-1}D_+^{-1}\per
  \nonumber
\end{equation}
From
\begin{equation}
  G_1\doteq
  D_+^{-1}-D_+^{-1}T_{++}D_+^{-1}\co\nonumber
\end{equation}
one has
\begin{equation}
  T_{++}=eB\frac{1}{1+(D_+^{-1}+D_-^{-1})eB}\per
  \nonumber
\end{equation}
This agrees indeed with
\begin{equation}
  dTd=d\frac{1}{1-elD_M^{-1}}el
  d\per\nonumber
\end{equation}

%%% Local Variables: 
%%% mode: latex
%%% TeX-master: "~/diss/tex/diss"
%%% End: 
\chapter{Determinants}
%--------------------------------
\label{app:det}
Consider the generating functional of Green's functions
\begin{equation}
  Z[j,j^*,J] = \frac{\int[d\hf][d\hf^*][d\lf]e^{i\int\La + j^*\hf + \hf^*j
  +J\lf}} {\int [d\hf][d\hf^*][d\lf]e^{i\int\La}}
\end{equation}
for the toy model
\begin{align}
  \La &= \La^0 + \La_\lf^0 + e\hf^*\hf \lf \notag \\
  \La^0 &= \partial_\mu\hf^*\partial^\mu\hf-M^2\hf^*\hf \notag \\
  \La_\lf^0 &= \frac{1}{2}\partial_\mu \lf\partial^\mu\lf + \frac{m^2}{2}\lf^2
\end{align}
studied in chapter~\ref{chap:toy-model}. Performing the integration over $\hf$
we get
\begin{equation}
  \label{eq:Z-hf-int}
  Z[j,j^*,J] = \frac{\int[d\lf](\det D_M^{-1}D_e)^{-1}e^{i\int  \mathcal
  L_\lf^0 + j^*D_e^{-1}j+J\lf}}{\int [d\lf](\det D_M^{-1}D_e)^{-1}e^{i\int
  \mathcal L_\lf^0}}.
\end{equation}
The operator $D_e$ was introduced in section~\ref{sec:external-field}
\begin{equation}
  D_e \doteq D_M-e\lf
\end{equation}
and $D_M = \Box + M^2$. The factors of $\det D_M$ were added to
eq.~\eqref{eq:Z-hf-int} for later convenience.

In this appendix, we first show that
\begin{equation}
  \label{eq:det-identity}
  \det D_M^{-1}D_e = \det D_+^{-1}\mathcal D_+ D_-^{-1}C,
\end{equation}
with the symbols
\begin{align}
  \mathcal D_+ &= D_+ + eB - e^2BC^{-1}B \\
  D_\pm &= \pm i\partial_t-\sqrt{M^2-\Delta} \\
  C &= D_- + eB \\
  B &= d\lf d
\end{align}
introduced in section~\ref{sec:external-field}. This decomposition allows us
to prove that Z may be written in the form
\begin{equation}
  \label{eq:Z-equiv}
  Z[j,j^*,J] = \frac{\int [d\lf](\det D_+^{-1}\mathcal D_+)^{-1}e^{i\int
  \mathcal L_\lf^0 + j^*D_e^{-1}j+J\lf}}{\int [d\lf](\det D_+^{-1}\mathcal
  D_+)^{-1}e^{i\int \mathcal L_\lf^0}}.
\end{equation}

The determinants are ill-defined as long as we don't specify how to deal with
the UV divergences inherent in their definitions.  In the following we use
dimensional regularization and work in $D\neq 4$ dimensions to render all
expressions finite. The statements derived here are then valid to all orders
in perturbation theory. Actual renormalization to one loop is performed in
appendix~\ref{app:1-loop-renormalization}.

In the following, we use the propagators $\Delta_M$ and $\Delta_\pm$ defined
in appendix~\ref{app:KG} as representations of the operators $D_M^{-1}$ and
$D_\pm^{-1}$, respectively.

\begin{figure}[thbp]
  \begin{center}
    \leavevmode \includegraphics[width=8cm]{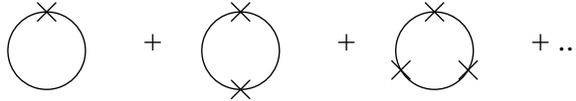}
    \caption{A graphical representation of the r.h.s. of
      eq. \eqref{eq:ln-det-De}. The line stands for a propagator $\Delta_M$
      and the cross for a light field $\lf$.}
    \label{fig:det-expansion}
  \end{center}
\end{figure}

Consider first the l.h.s. of eq.~\eqref{eq:det-identity}
\begin{displaymath}
  \det D_M^{-1}D_e = \det(1-D_M^{-1}e\lf) = e^{\Tr\ln(1-D_M^{-1}e\lf)}.
\end{displaymath}
Expanding the logarithm we can write
\begin{multline}
  \label{eq:ln-det-De}
  \Tr\ln(1-D_M^{-1}e\lf) = -\sum_{n=1}^{\infty}\frac{1}{n}\Tr(
  D_M^{-1}e\lf)^n \\
  = -\sum_{n=1}^\infty \frac{1}{n} \int d^Dx_1\dots d^Dx_n
  \Delta_M(x_1-x_2)e\lf(x_2) \dots \Delta_M(x_n-x_1)e\lf(x_1).
\end{multline}
The $n^{\text th}$ term of this sum is a loop formed by connecting $n$ fields
$\lf$ by as many propagators $\Delta_M$ (see figure \ref{fig:det-expansion}).
Using the identity $C^{-1}eB = 1-C^{-1}D_-$, which follows directly from the
definition of $C$, we can cast the r.h.s. of eq. \eqref{eq:det-identity} into
the form
\begin{displaymath}
  \det D_+^{-1}\mathcal D_+ D_-^{-1}C =
  \det\left(1+\left(D_+^{-1}+D_-^{-1}\right)eB\right). 
\end{displaymath}
Proceeding as before, we find
\begin{multline}
  \label{eq:ln-det-D-p}
  \Tr\ln\left(1+\left(D_+^{-1}+D_-^{-1}\right)ed\lf d\right) =
  \sum_{n=1}^\infty \frac{(-1)^{n+1}}{n} \int d^Dx_1\dots d^Dx_n \\
  d_{x_1}(\Delta_+(x_1-x_2)+\Delta_-(x_1-x_2))d_{x_2}e\lf(x_2) \\ \dots
  d_{x_n}(\Delta_+(x_n-x_1)+\Delta_-(x_n-x_1))d_{x_1}e\lf(x_1),
\end{multline}
where the subscript of $d$ indicates on which variable it acts. Due to the
operator relation
\begin{equation}
  \label{eq:DM-decomp}
  D_M^{-1} = -d(D_+^{-1} + D_-^{-1})d
\end{equation}
derived in appendix~\ref{app:KG} this is indeed equal to the l.h.s. of
eq.~\eqref{eq:det-identity}.

Formally, we can write eq.\eqref{eq:det-identity} as
\begin{equation}
  \label{eq:det-identity-2}
  \det D_M^{-1}D_e = \det D_+^{-1}\mathcal D_+ \det D_-^{-1}C.
\end{equation}
Taking the logarithm, we find to first order in $e$ the tadpole term
\begin{displaymath}
  \Delta_M^D(0)\int d^Dx \lf(x) = - \left(\left. d^2\Delta_+^D(x)\right|_{x=0}
  + \left.d^2\Delta_-^D(x)\right|_{x=0}\right)\int d^Dx\lf(x).
\end{displaymath}
The explicit expressions of the terms on the r.h.s. are
\begin{equation}
  \left.d^2\Delta_\pm^D\right|_{x=0} = - \int \frac{d^D
  p}{(2\pi)^D}\frac{1}{2\omega(\mathbf p)(\omega(\mathbf p)\mp p^0 -
  i\epsilon)}. 
\end{equation}
In standard dimensional regularization, where one writes $d^Dp =
dp^0d^{D-1}\mathbf p$ and integrates over $p^0$ separately, these are not well
defined because the integrand falls off only like $1/p^0$ for large $p^0$. One
may use what is called split dimensional regularization~(\cite{split-d}) where
one introduces two independent regulators $\sigma$ and $D$ according to
\begin{equation}
  d^D p = d^\sigma p^0 d^{D-\sigma}\mathbf p.
\end{equation}
In this scheme we find
\begin{equation}
  \left. d^2\Delta_+^D(x)\right|_{x=0}
  + \left.d^2\Delta_-^D(x)\right|_{x=0} = -e^{i\sigma\frac{\pi}{2}}
  \frac{M^{D-2}}{(4\pi)^\frac{D}{2}} \Gamma(1-\frac{D}{2})
\end{equation}
and one can check that this is indeed equal to $-\Delta_M(0)$, evaluated with
the same prescription.

This subtlety only occurs in the tadpole. Every other graph has an integrand
that falls off at least like $1/(p^0)^2$ and split dimensional regularization
coincides with the standard dimensional regularization.
Eq.~\eqref{eq:det-identity-2} is therefore true within this special
regularization scheme.

Let us give an intuitive understanding of this decomposition. A loop
containing $n$ propagators can be written as a sum of $2^n$ terms by
decomposing $\Delta_M$ into $\Delta_\pm$ as in eq.  \eqref{eq:DM-decomp}. One
of these terms will exclusively contain anti-particle propagators $\Delta_-$
and all of these graphs are collected in the expression $\det D_-^{-1}C$.
Therefore, eq.  \eqref{eq:det-identity-2} can be interpreted as the separation
of the contribution of the pure anti-particle sector to loops formed by the
heavy field.

\begin{figure}[thbp]
  \begin{center}
    \leavevmode \includegraphics[width=5cm]{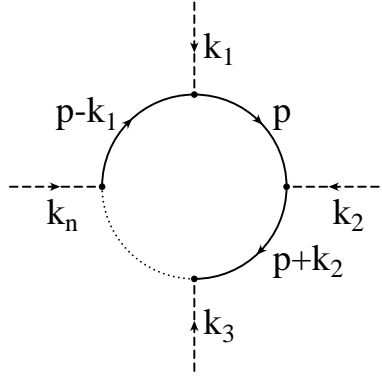}
    \caption{A typical contribution of a loop formed exclusively with
      anti-particle propagators (solid lines). It is connected to the rest of
      the diagram only by propagators of the light field (dashed lines).}
    \label{fig:det-C-contrib}
  \end{center}
\end{figure}

To prove eq. \eqref{eq:Z-equiv}, we show that $\det D_-^{-1}C$ does not
contribute to any Green's functions contained in Z. The explicit expression
for this determinant is
\begin{displaymath}
  \begin{split}
    \det D_-^{-1}C = \exp\Bigl\{\sum_{n=1}^\infty \frac{(-1)^{n+1}}{n} \int
    d^Dx_1 \dots d^Dx_n \Delta_-(x_1-x_2)eB(x_2)\dots \Bigr. \\ \Bigl. \times
    \Delta_-(x_n-x_1)eB(x_1) \Bigr\}.
  \end{split}
\end{displaymath}
Upon performance of the $\lf$-integration in eq. \eqref{eq:Z-hf-int} it will
produce sub-graphs of the type shown in figure \ref{fig:det-C-contrib}.
Because $\Delta_-(p^0,\vec p^2)$ contains only one pole in $p^0$, all poles of
the integrand of such a loop lie in the same half-plane. We can close the
contour of the integration in the other half-plane and find that the entire
integral vanishes, irrespective of the rest of the diagram that this loop is
part of. Therefore, we can drop this determinant in the expression for $Z$
{\em without changing any Green's functions}, which completes the proof of eq.
\eqref{eq:Z-equiv}.

Let us return to the tadpole contribution discussed above. We may separate it
by defining
\begin{equation}
  \label{eq:tadpole-rem-1}
  \delta \doteq (\det D_M^{-1}D_e)^{-1} e^{-e\Delta_M(0)\int d^Dx l(x)}
\end{equation}
Furthermore, let us add a term to the Lagrangian
\begin{equation}
  \bar \La = \La - e\frac{1}{i}\Delta_M(0)\lf. 
\end{equation}
The corresponding generating functional
\begin{equation}
  \label{eq:tadpole-rem-2}
  \bar Z[j,j^*,J] = \frac{\int [d\lf]\delta e^{i\int\La_\lf^0 + j^*D_e^{-1}j +
  Jl}} 
  {\int [dl]\delta e^{i\int\La_\lf^0}}
\end{equation}
is identical to $Z$, except that it does not contain any one-loop tadpole
contributions. The additional term in the definition of $\bar\La$ can be
viewed as a 1-loop counter term. We have thus shown that renormalization can be
done in such a way that the tadpole is removed from any Green's function (see
also appendix~\ref{app:1-loop-renormalization}).

%%% Local Variables: 
%%% mode: latex
%%% TeX-master: "~/diss/tex/diss"
%%% End: 
\chapter{1-Loop Renormalization}
%===============================
\newcommand{\pa}{\phi_1}
\newcommand{\pb}{\phi_2}
\newcommand{\pc}{\phi_3}
\newcommand{\pT}{\phi^{\text{T}}}
\newcommand{\jT}{j^{\text{T}}}
\label{app:1-loop-renormalization}
\section{Relativistic Theory}
%----------------------------
We consider the tree-level Lagrangian
\begin{equation}
  \La = -\hf^*D_M\hf-\frac{1}{2}\lf D_m\lf + e\hf^*\hf\lf
  +j^*\hf + \hf^*j + J\lf,
\end{equation}
where $D_M=\Box+M^2$ and $D_m=\Box+m^2$. It is convenient to replace the
complex field $\hf$ by two real fields $\pa$, $\pb$ and the source $j$ by two
real sources $j_1$, $j_2$ through
\begin{align}
  \hf &= \frac{1}{\sqrt{2}}(\pa + i\pb) \notag \\
  j &= \frac{1}{\sqrt{2}}(j_1 + ij_2).
\end{align}
Renaming $\lf\equiv\pc$, $J\equiv j_3$, we can collect the fields and sources
in three-di\-men\-sion\-al vectors
\begin{align}
  \pT &= (\pa, \pb, \pc) \notag \\
  \jT &= (j_1,j_2,j_3).
\end{align}
The Lagrangian then reads
\begin{equation}
  \La = -\frac{1}{2}\pT D_0\phi + \frac{e}{2}(\pa^2+\pb^2)\pc + \jT\phi,
\end{equation}
with
\begin{equation}
  D_0 = 
  \begin{pmatrix} 
    D_M & 0 & 0 \\
    0 & D_M & 0 \\
    0 & 0 & D_m
  \end{pmatrix}.
\end{equation}
In units where $\hbar$ is explicit, the generating functional $W$ of connected
Green's functions is defined by
\begin{equation}
  e^{\frac{i}{\hbar}W[j;\hbar]} = \frac{1}{\mathcal Z} \int [d\phi]
  e^{\frac{1}{\hbar} S[j]},
\end{equation}
where
\begin{align}
  \mathcal Z &= \int [d\phi] e^{\frac{i}{\hbar} S[0]} \notag \\
  S[j] &= \int d^dx \La(x;j).
\end{align}
We use dimensional regularization to give a meaning to the path integral and
want to construct the counter term Lagrangian
\begin{equation}
  \La_{ct} = \hbar\La^{(1)} + O(\hbar^2)
\end{equation}
that absorbs the divergences in $d=4$. The expansion
of $W$ in powers of $\hbar$ is equivalent to an expansion in the number of
loops, so that $W_0$ and $W_1$ defined by
\begin{equation}
  W[j;\hbar] = W_0[j] + \hbar W_1[j] + O(\hbar^2)
\end{equation}
generate tree- and one-loop graphs, respectively. This expansion is obtained
by writing $\phi$ as fluctuation around the solution $\bar\phi$ of
the equations of motion
\begin{align}
  D_M\bar\pa - e\bar\pa\bar\pc - j_1 &= 0 \notag \\
  D_M\bar\pb - e\bar\pb\bar\pc - j_2 &= 0 \notag \\
  D_m\bar\pc -\frac{e}{2}(\bar\pa^2 + \bar\pb^2)\bar\pc - j_3 &= 0.
\end{align}
Setting $\phi = \bar\phi + \hbar^{1/2}\eta$ and keeping only terms of
$O(\hbar)$ we find
\begin{align}
  W_0 &= \int d^dx \bar\La(x) \\
  \label{eq:W1}
  W_1 &= \frac{i}{2}\ln\frac{\det D}{\det D_0} + \int d^dx\bar\La^{(1)}(x),
\end{align}
where
\begin{equation}
  D = D_0 -e 
  \begin{pmatrix}
    \bar\pc & 0 & \bar\pa \\
    0 & \bar\pc & \bar\pb \\
    \bar\pa & \bar\pb & 0
  \end{pmatrix}
\end{equation}
and barred quantities are evaluated at $\phi=\bar\phi$. Applying the heat
kernel technique, the contributions to $W_1$ that diverge in $d=4$ can be
isolated. The result is
\begin{align}
  \label{eq:W1div}
  W_1 = &\; \frac{e^2}{2} \Delta_1 \int d^dx \left( \bar\pa^2(x) + \bar\pb^2(x)
  \right) + \frac{e^2}{2}\Delta_2 \int d^dx
  \bar\pc^2(x) \notag \\
  & + e \Delta_3 \int d^dx\bar\pc(x) + \text{finite}(d\rightarrow 4),
\end{align}
with
\begin{align}
  \Delta_1 &= \frac{1}{2} \frac{\Gamma(-\omega)}{(4\pi)^{2+\omega}} 
  \left(M^{2\omega}+m^{2\omega}\right) \\
  \Delta_2 &= \frac{\Gamma(-\omega)}{(4\pi)^{2+\omega}} M^{2\omega} \\
  \Delta_3 &= \frac{\Gamma(-1-\omega)}{(4\pi)^{2+\omega}} M^{2(\omega+1)}
\end{align}
and $\omega=(d-4)/2$. We introduce the renormalization scale $\mu$ with the
object
\begin{align}
  \hat L &= \left(\frac{M}{\mu}\right)^{2\omega} \frac{\mu^{2\omega}}{32\pi^2}
  \frac{\Gamma(-1-\omega)}{(4\pi)^\omega} \notag \\
  &= L(\mu) + \frac{\mu^{2\omega}}{32\pi^2}
  \left(\ln\frac{M^2}{\mu^2} -1 \right) + a(\omega,\frac{M}{\mu}) \\
  L(\mu) &= \frac{\mu^{2\omega}}{32\pi^2}\left(\frac{1}{\omega} - \Gamma'(1) -
  \ln 4\pi \right).
\end{align}
The function $a$ vanishes in the limit $\omega\rightarrow 0$ and is not needed
explicitly. $\hat L$ is independent of $\mu$
\begin{equation}
  \mu\frac{\partial}{\partial\mu} \hat L = 0
\end{equation}
and so are
\begin{align}
  \Delta_1 &= -2\left[ L(\mu) + \frac{\mu^{2\omega}}{32\pi^2}\left\{
      \ln\frac{M^2}{\mu^2} + \frac{1}{2}\ln\frac{m^2}{M^2}\right\} 
    + b(\omega,\frac{M}{\mu},\frac{m}{\mu}) \right] \\
  \Delta_2 &= -2\left(L(\mu) +
      \frac{\mu^{2\omega}}{32\pi^2}\ln\frac{M^2}{\mu^2} + 
    c(\omega,\frac{M}{\mu})\right)  \\
  \Delta_3 &= 2M^2\hat L
\end{align}
Like $a$, the functions $b$ and $c$ vanish for $\omega\rightarrow 0$. In order
to cancel these divergences, we need a counter term Lagrangian of the form
\begin{equation}
  \La^{(1)} = -\frac{e^2}{2}c_1\left(\pa^2+\pb^2\right) -
  \frac{e^2}{2}c_2\pc^2 - c_3 eM^2\pc,
\end{equation}
The dimensionless constants $c_n$ can be chosen to be independent of $\mu$ and
in the $\overline{MS}$ scheme we set
\begin{equation}
  c_n = c_{n}^r(\mu,\omega) + \Gamma_n L(\mu).
\end{equation}
The renormalized couplings $c_n^r$ are finite and depend on the scale
according to the renormalization group equations
\begin{equation}
  \mu\frac{\partial}{\partial\mu} c_n^r(\mu,\omega) =
  -2\omega\Gamma_n L(\mu).
\end{equation}
From eq.~\eqref{eq:W1div} we can read off
\begin{align}
  \Gamma_{1,2} = -2.
\end{align}
The term $\Delta_3$ plays a special role. In appendix~\ref{app:det} it was
identified with the loop of the tadpole graph
\begin{displaymath}
  \includegraphics[scale=0.7]{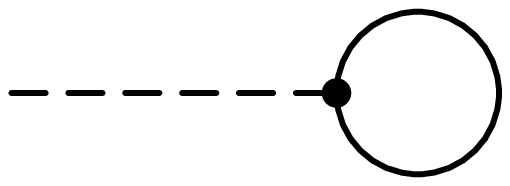}
\end{displaymath}
which is simply the Fourier transform $\Delta_M(0)$ of the heavy propagator at
zero momentum. In fact, we have $\Delta_3 = -i\Delta_M(0)$. Now, in that
appendix it was shown that by adding the term $ie\Delta_M(0)\lf$ to the
Lagrangian, the tadpole is removed from all the Green's functions (see
eqns.~\eqref{eq:tadpole-rem-1} through~\eqref{eq:tadpole-rem-2}). We therefore
chose
\begin{equation}
  c_3 = 2\hat L.
\end{equation}

Physical quantities can be expressed in terms of the scale-independent and
finite couplings 
\begin{equation} 
  \bar c_n = -c_n^r(\mu,0) + \frac{\Gamma_n}{32\pi^2}\ln\frac{M^2}{\mu^2}.
\end{equation}
They are determined through the condition that the parameters $M$ and $m$
should coincide with the physical masses $\Mp$ and $m_{\text{p}}$. The
explicit expressions are not needed here.

Finally, we may go back to the original fields and find that
\begin{equation}
  \La(\hf,\hf^*,\lf) - c_1 e^2 \hf^*\hf - c_2\frac{e^2}{2}\lf^2 -
  c_3 eM^2\lf 
\end{equation}
gives finite results in $d=4$ at 1-loop level. 

\section{Effective Non-local Theory}
%------------------------------------
In appendix~\ref{app:det} it is shown that the non-local Lagrangian
\begin{equation}
  \La = \La_+ + j^*\hfp + \hfp^* j + J\lf
\end{equation}
constructed in section~\ref{sec:p-a-sector} contains the same loops as the
full theory. Therefore, the only divergent graphs to one loop are the
self-energies and the vacuum polarization of the light field. 

Consider first the two-point function of the light field. We know that it is
identical in both theories (this is evident by comparing the
expressions~\eqref{eq:Z-integrated} and~\eqref{eq:Z-hp-integrated} of the
generating functionals) and so must be the counter terms.  

By comparing the two-point functions of the heavy field in
eqns. \eqref{eq:G2-orig} and \eqref{eq:G2-eff} we find that
$-e^2c_1(d\hfp^*)(d\hfp)$ is the counter term needed to render the self-energy
of $\hf$ in the effective theory finite. 

The vacuum expectation value of the light field is given by
\begin{equation}
  v = \int d^4 x e^{ipx}\eval{0}{\lf(x)}{0} = (2\pi)^4\delta^4(p)\Delta_m(p)
  ie[d^2\Delta_+](0) 
\end{equation}
where as in the relativistic theory we have
\begin{equation}
  v = (2\pi)^4\delta^4(p)\Delta_m(p) ie\Delta_M(0).
\end{equation}
The quantity $[d^2\Delta_+](0)$ is evaluated in split dimensional
regularization discussed in appendix~\ref{app:det}
\begin{align}
  [d^2\Delta_+](0) &= -\frac{1}{2}e^{i\sigma\frac{\pi}{2}}\Delta_3 \nonumber\\
  &= -M^2 e^{i\sigma\frac{\pi}{2}}\hat L.
\end{align}
The appropriate counter term is therefore $-\tilde c_3 eM^2l$ with 
\begin{equation}
  \tilde c_3 \doteq e^{i\sigma\frac{\pi}{2}}\hat L.
\end{equation}

Putting everything together, the effective Lagrangian that is finite at 1-loop
is given by
\begin{equation}
  \La_+ - c_1e^2 (d\hfp^*)(d\hfp) 
  - c_2\frac{e^2}{2}\lf^2 - \tilde c_3 eM^2l.
\end{equation}

%%% Local Variables: 
%%% mode: latex
%%% TeX-master: "~/diss/tex/diss"
%%% End: 
\end{appendix}
\bibliographystyle{hunsrt}
\bibliography{misc,hqet,nrqcd-qed,hbchpt,pipi}
\markboth{BIBLIOGRAPHY}{}
\end{document}